\documentstyle[preprint,aps,tighten,epsf,prb]{revtex} 

\begin{document} 
 
 
\title{Coulomb blockade conductance peak fluctuations in quantum dots
and the independent particle model}
 
\author{Ra\'ul O. Vallejos$^1$, Caio H. Lewenkopf$^1$,
        and Eduardo R. Mucciolo$^2$} 
 
\address{$^1$ 
     Instituto de F\'{\i}sica,  
     Universidade do Estado do Rio de Janeiro, \\ 
     R. S\~ao Francisco Xavier 524, 20559-900 Rio de Janeiro, Brazil\\ 
         $^2$ 
     Departamento de F\'{\i}sica,  
     Pontif\'{\i}cia Universidade Cat\'olica do Rio de Janeiro,\\ 
     Caixa Postal 38071, 22452-970 Rio de Janeiro, Brazil} 
 
\date{\today} 
 
\maketitle 
 

\begin{abstract}
We study the combined effect of finite temperature, underlying
classical dynamics, and deformations on the statistical properties of
Coulomb blockade conductance peaks in quantum dots. These effects are
considered in the context of the single-particle plus
constant-interaction theory of the Coulomb blockade. We present
numerical studies of two chaotic models, representative of different
mean-field potentials: a parametric random Hamiltonian and the smooth
stadium. In addition, we study conductance fluctuations for different
integrable confining potentials. For temperatures smaller than the
mean level spacing, our results indicate that the peak height
distribution is nearly always in good agreement with the available
experimental data, irrespective of the confining potential (integrable
or chaotic). We find that the peak bunching effect seen in the
experiments is reproduced in the theoretical models under certain
special conditions. Although the independent particle model fails, in
general, to explain quantitatively the short-range part of the peak
height correlations observed experimentally, we argue that it allows
for an understanding of the long-range part.
\end{abstract} 

 
 
 
 
\narrowtext 


\section{Introduction}
\label{sec:introduction}

 
Since the first electronic transport measurements in semiconductor
quantum dots in the Coulomb blockade regime were made,\cite{Kastner92}
a significant amount of data has been accumulated. The appropriate
theoretical model used in understanding this data depends on the size
of the electronic island. Two situations arise. When dealing with
small or very small systems, where the number of electrons $N$ in the
Coulomb island is of the order of 10 (or a maximum of $10^2$), the
self-consistent approach is successful.\cite{smalldotrefs} In this
regime the environment generates a smooth, nearly parabolic potential
and most of the quantum dot properties can be understood in terms of
an unsophisticated self-consistent electronic structure
calculation. On the other hand, the most adequate theoretical
description for data taken from larger systems, with $N$ of order
$10^3$ (or at least larger than $10^2$), is statistical. In
distinction to the first case, for large dots the confining potential
is no longer expected to be parabolic up to energies close to the
Fermi surface. In fact, in large gated structures, the dot geometrical
shape can be substantially modified by adjusting voltages. Within the
single-particle picture, it is believed that the complexity of the
confining potential yields a chaotic motion in the classical limit,
thus justifying a quantum mechanical modeling based on a statistical
theory.\cite{Beenakker97,Guhr98} Further arguments for such approach
are provided by the influence of weak disorder and the fact (not
always justifiable) that many-body correlations should bring only
small corrections to the usual single-particle description of
transport.

The single-particle statistical
theory\cite{Beenakker91,Jalabert92,Prigodin93} is widely accepted
because of its early success in predicting the distribution of peak
heights in the single-channel tunneling regime, later experimentally
confirmed.\cite{Chang96,Folk96} In the regime where the mean level
spacing $\Delta$ is much larger than $k_BT$, the predicted
distribution of peak heights is strongly non-Gaussian and clearly
peaked towards zero values. While the single-particle plus constant
interaction theory captures most qualitative aspects observed in the
experiments, some quantitative aspects defy comprehension. Perhaps the
most celebrated one is the absence of Wigner-Dyson fluctuations in the
conductance peak spacings.\cite{Sivan96,Simmel97,Patel97}

There is also an important and rather prominent feature of the
experimental data which is less discussed in the literature and
remains unexplained: the notable clustering of peaks with similar
heights. That is, in most experiments (an important exception is
Ref.~\onlinecite{Chang96}), by sweeping the gate voltage, large and
small peaks tend to appear in bunches, indicating the existence of
peak to peak correlation. This is at odds with the statistical
single-particle theory at low temperatures, which associates to each
peak an independent single-particle eigenstate. A recent experiment
\cite{Patel98} showed yet another puzzling feature: the correlation
tends to increase with temperature and the slope depends on the dot
area (for a small dot it saturated at $k_BT \approx \Delta/2$.)

Any successful theoretical model describing all these issues has to
take into account that each conductance peak corresponds to a system
containing a different number of particles. As electrons are added to
the quantum dot the gate voltage configuration varies and the
confining potential changes accordingly.\cite{Vallejos98} Furthermore,
and probably even more important, the additional particles change the
electron mean field. Although not the original idea addressed in
Ref.~\onlinecite{Vallejos98}, we can parameterize the total change in
the single-particle energies {\it and} wave functions by moving from
one peak to the next through a parametric variation $\delta X$ which
takes into account changes of the external and mean field
potentials. We can call it the mean field ``deformation''. Although
difficult to quantify, $\delta X$ is obviously nonzero.
 
In this paper we concentrate on the statistical measures of the
conductance peak heights, namely their distribution and
autocorrelation function. We use the constant interaction plus
independent particle model to discuss the role of ``deformations'' and
temperature. We show that for $k_BT \le \Delta$ the theoretical peak
height distribution agrees with the experimental data for systems with
chaotic underlying dynamics, irrespective of the deformation
parameter. We also show a very reasonable agreement between chaotic
and {\sl some} integrable models for $k_B T > 0.25 \Delta$. As for the
peak height autocorrelation function, we could not find any robust and
generic mechanism to explain the strong peak to peak correlations seen
in some experiments. We were only able to find such correlations for
special choices of lead positions.

The structure of this paper is as follows. In section
\ref{sec:thetheory} we begin with a short review of the constant
interaction model, which is used throughout this work, centered on a
discussion about finite temperature corrections and parametric
variations. In section \ref{sec:rmt} we study the combined effect of
temperature and deformations in a random matrix model. In section
\ref{sec:dynamical} we investigate dynamical mechanisms to explain the
observed correlations between peak heights. The proposed mechanisms
are examined by studying a suitable dynamical model. The influence of
data set sizes on the evaluation of peak height correlations is
critically discussed in section \ref{sec:finitedataset}. Having
analyzed only chaotic dynamical systems, we investigate in section
\ref{sec:integravel} whether the experiments carried out so far indeed
observed fingerprints of chaos. Insight on this question is obtained
by studying the peak height distribution for a few integrable
models. Finally, we summarize our conclusions in section
\ref{sec:conclusions}.

 
\section{The single-particle plus constant interaction model for 
quantum dots}
\label{sec:thetheory} 
 
 
The theoretical description used in this study is based on the
so-called independent particle plus ``constant interaction" (CI)
model,\cite{Kastner92} where the electron-electron interaction is
taken into account through a fixed (capacitive) charging energy term
in the Hamiltonian:
\begin{equation}
\label{eq:omodelo}
H = \sum_{i=1}^N \left[ \frac{\bbox{p}_i^2}{2m^\ast} + V(\bbox{r}_i)
    \right] + \frac{U}{2} N(N-1) + e\eta V_g N,
\end{equation}
where $N$ is the total number of electrons with effective mass
$m^\ast$ in the dot, $U$ is the charging energy, and $V_g$ is some
tunable gate voltage. The coefficient $\eta$ is a function of the
capacitance matrix elements of the dot.\cite{Glazman89} The single
particle potential in Eq.~(\ref{eq:omodelo}) accounts only for the
electrostatic confinement and background disorder.

When the dot is in equilibrium with the reservoirs represented by the
leads, its total energy is minimal for a certain value of $N$. The
precise value of $N$ depends on the dot confining geometry, chemical
potential (Fermi energy) in the reservoirs, and gate voltage
$V_g$. The electronic many-body ground state wave function is given by
the Slater determinant of all bounded single-particle eigenstates with
energies smaller than the Fermi energy. In its crudest version, the
single-particle states in the CI model are determined by the bare
electron confining potential alone. It has been argued
\cite{Berkovits98,Levit99,Bonci99,Walker99,Cohen99} that many-body
effects are important and one should instead get the single-particle
states from a mean-field approximation, such as Hartree-Fock. We will
not attempt to answer this question in this work, but rather
comprehensively study most features that may be traced to one-body
effects.

Both single-particle eigenenergies and eigenfunctions influence the
electronic transport through the quantum dot. The wave functions, in
particular, enter into the standard formulation\cite{Beenakker91}
through the partial decay widths $\Gamma^{r,l}$ which quantify the
connectivity between the eigenstate in the dot and the electronic
states in the leads ($r$ and $l$ stand for right and left leads,
respectively). In the regime of thermally broadened resonances and
large charging energy, $\Gamma^{r,l} \ll k_BT \ll U$, linear response
theory yields the following expression for the the two-point
conductance of a dot (weakly) coupled to reservoirs\cite{Beenakker91}
\begin{eqnarray} 
G & = & \frac{e^2}{h k_BT} \sum_{p=1}^{\infty} \frac{\Gamma_p^l
\Gamma_p^r}{\Gamma_p^l + \Gamma_p^r}
F_{\mbox{\scriptsize{eq}}}(\varepsilon_p|N) \nonumber \\ & & \times
\left[ 1 - f ( \delta E - \varepsilon_N + \varepsilon_p ) \right]
\nonumber \\ & & \times f \biglb( \delta E -TS(N)+TS(N-1) \bigrb),
\label{eq:conductance}
\end{eqnarray} 
with $\delta E \equiv \varepsilon_N + U(N-1) - e \eta V_g -
\varepsilon_F$. The sum runs over all single-particle states $p$ with
energies $\varepsilon_p$ and partial decay widths $\Gamma_p^r$ and
$\Gamma_p^l$. The latter describe the electron tunneling out of the
dot and into the reservoirs, or vice-versa. Thus, the total tunneling
rate is $h \Gamma_p = h (\Gamma_p^l +
\Gamma_p^r)$. $F_{\mbox{\scriptsize{eq}}}(\varepsilon_p|N)$ is the
canonical occupation probability of the level $p$ for a dot containing
$N$ electrons in thermal equilibrium. $f$ is the Fermi-Dirac
distribution, $f(\varepsilon)=1/[1+\exp(\varepsilon/k_BT)]$, $S$ is
the entropy of the dot, and $\varepsilon_F$ is the Fermi energy in the
reservoirs. We will specialize our discussion to spin split
(nondegenerate) energy levels.

In experiments, sequences of conductance peaks are observed by varying
the external gate voltage $V_g$, thus changing the electrostatic
energy in the dot. Within the CI model, in the single-level tunneling
regime, where $\Gamma^{r,l} \ll k_B T \ll \Delta$, and as the gate
potential is swept, conductance peaks occur whenever the resonance
condition
\begin{equation}
E(N+1) - E(N) = \varepsilon_F
\label{eq:resonancecond}
\end{equation} 
is satisfied.\cite{Beenakker91} Here $E(N)$ is the ground state energy
of the dot containing $N$ electrons. Within the constant interaction
model, Eq.~(\ref{eq:resonancecond}) can be rewritten as
\begin{equation} 
\label{eq:resonance} 
\varepsilon_N + (N -1)U - e \eta V_g = \varepsilon_F,
\end{equation} 
i.e., $\delta E = 0$. We remark that, although correct at the level of
the Hartree-Fock approximation (with $N\gg 1$), it is still a matter
of debate in the literature whether Eq.~(\ref{eq:resonance}) is
generally valid.
\cite{Berkovits98,Levit99,Bonci99,Walker99,Cohen99,Blanter97,Berkovits97}
The existence of many-body correlation effects may lead to a breakdown
of Koopmans' theorem and a different resonance condition.
Furthermore, the strong assumption of a constant $U$ prevents an
adequate description of fluctuations in peak position. Nevertheless,
these restrictions shall not affect our discussion on the statistics
of peak heights in the single-particle approximation.

In order to make the notation more compact, let us rescale the
conductance to a dimensionless form,
\begin{equation}
G \equiv \frac{e^2}{h} \frac{\langle \Gamma \rangle}{4k_BT} \,g .
\end{equation}
Moreover, for the sake of simplicity, we consider that the average
decay widths (taken over many resonances) are equal: $\langle
\Gamma^{r}\rangle = \langle \Gamma^{l}\rangle = \langle
\Gamma\rangle$. In the temperature regime where $\langle \Gamma
\rangle \ll k_B T \ll \Delta$, the resonance condition, combined with
Eq.~(\ref{eq:conductance}), results in the well-known expression for
the {\it single-level} conductance peak height,
\begin{equation}
\alpha_p \equiv g^{\mbox{\scriptsize{max}}}_p =  \frac{1}{\langle \Gamma
         \rangle} \frac{\Gamma_p^l \Gamma_p^r}{\Gamma_p^l +
         \Gamma_p^r}.
\end{equation}
Assuming that the electronic dynamics inside the dot is classically
chaotic, Jalabert and collaborators \cite{Jalabert92} used a random
matrix model to calculate the distribution of partial decay widths in
this temperature regime, in the presence ($\beta=2$) or not
($\beta=1$) of an applied magnetic field. They obtained
\begin{equation}
\label{eq:Pgamma}
P_\beta(\Gamma^{r(l)}) = \frac{C_\beta}{2 \Gamma^{r(l)} }
      \left(\frac{\beta\Gamma^{r(l)}}{2\langle\Gamma^{r(l)}\rangle}
      \right)^{\beta/2-1} \exp\left( \frac{\beta\Gamma^{r(l)}}
      {2\langle\Gamma^{r(l)}\rangle} \right),
\end{equation}
where $C_\beta$ is a normalization constant that can be expressed in
terms of the gamma function: $C_\beta=\beta/\Gamma(\beta/2)$. By
further assuming that the contacts are sufficiently far from each
other so that $\Gamma^r$ and $\Gamma^l$ are uncorrelated, expressions
were derived for the distribution of dimensionless peak heights,
namely,
\begin{equation}
\label{eq:PGOEg}
P_{\beta=1}(\alpha) = \sqrt{\frac{2}{\pi \alpha}}\ e^{-2\alpha}
\end{equation}
and
\begin{equation}
\label{eq:PGUEg}
P_{\beta=2}(\alpha) = 4\alpha\Big[K_0(2\alpha) + K_1(2\alpha)\Big]
e^{-2\alpha},
\end{equation}
where $K_0(x)$ and $K_1(x)$ are modified Bessel functions. These
predictions were later experimentally confirmed by two independent
experiments.\cite{Chang96,Folk96} Equations~(\ref{eq:PGOEg}) and
(\ref{eq:PGUEg}) will serve us as benchmarks in later sections.

As mentioned in the introduction, the early success of the random
matrix hypothesis\cite{Jalabert92,Prigodin93} in explaining
statistical properties of conductance peaks in quantum dots seemed
somewhat inconsistent. The experimental confirmation of theoretical
prediction of the distribution $P(\alpha)$ contrasts with the
observation of strong correlations between successive peak heights in
most data sets.\cite{Folk96,Patel98} Recall that standard random
matrix theory (RMT)\cite{Mehta91} leads to uncorrelated
eigenfunctions, giving conductance peaks with a similar
property. However, the fact that the condition $k_B T\ll \Delta$ is
hardly achieved in the early experiments suggests that the discrepancy
could be related to a finite temperature effect. The inclusion of
finite temperature in RMT was carried out by Alhassid and
collaborators \cite{Alhassid98} by computing the canonical occupation
factors $F_{\mbox{\scriptsize{eq}}}(\varepsilon_p|N)$ in
Eq.~(\ref{eq:conductance}). Already for $k_B T \approx 0.3 \Delta$
they observed significant temperature corrections in $P(\alpha)$, as
compared to Eq.~(\ref{eq:PGUEg}). The corrections, however, fail to
quantitatively reproduce the available experimental
data,\cite{Patel98} and the discrepancy grows larger as temperatures
goes beyond $\Delta$. We will return to this point later.

The canonical average necessary to calculate
$F_{\mbox{\scriptsize{eq}}}(\varepsilon_p|N)$ can be taken either by
explicit summation of all possible single-particle configurations, as
recently done in Ref.~\onlinecite{Baltin98}, or in a quadrature
approximation for the canonical partition function, as in
Refs.~\onlinecite{Alhassid98,Ormand94}. We adopt the latter approach
in what follows. The procedure consists in expressing the canonical
partition function as
\begin{equation}
Z_N = \frac{ e^{-E^{\scriptsize{\mbox{tot}}}(N)/k_BT}} {L}
\sum_{j=1}^{L} \prod_{l=N-\frac{L}{2}}^{N+\frac{L}{2}} \left( 1 +
e^{i\phi_j \sigma_l - \frac{|\lambda - \varepsilon_l |}{k_BT}}
\right),
\label{eq:partitionfunction}
\end{equation}
where $\phi_j = 2\pi j/L$, $E^{\scriptsize{\mbox{tot}}}(N) =
\sum_{p=N-L/2}^{N+L/2} \varepsilon_p$, and
\begin{equation}
\sigma_l = \left\{ \begin{array}{rl} -1, &\qquad (l\le N) \\ 1,
&\qquad (l> N), \end{array} \right.
\end{equation}
with $L$ standing for the number of states taken into account (we are
interested in the case where $N\gg L\gg 1$). This form of organizing
the sum, involving only exponentially decaying terms, together with
the choice $\lambda = (\varepsilon_N + \varepsilon_{N+1})/2$,
guarantees fast convergence. The free energy and the entropy are
straightforwardly evaluated from Eq.~(\ref{eq:partitionfunction})
recalling that $F(N) = -k_B T \ln Z_N = E^{\scriptsize{\mbox{tot}}}(N)
- T S(N)$. With the quadrature approximation, the calculation of the
canonical occupation number is carried according to
\begin{eqnarray}
F_{\mbox{\scriptsize{eq}}}(\varepsilon_p|N) & = & \frac{ e^{-S(N)/k_B}
} {L} \sum_{j=1}^{L} \prod_{l=N-\frac{L}{2}}^{N+\frac{L}{2}} \left( 1
+ e^{i\phi_l \sigma_l - \frac{|\lambda - \varepsilon_l|}{k_BT}}
\right) \nonumber \\ & & \times \left( 1 + e^{ - i\phi_l \sigma_p +
\frac{|\lambda - \varepsilon_p|}{k_BT}} \right)^{-1}.
\end{eqnarray}

Since we are only interested in the statistics of conductance peak
maxima and not of spacings, we introduce in Eq.~(\ref{eq:conductance})
the resonance condition for $k_B T \ll \Delta$
[Eq.~(\ref{eq:resonance})]. In the limit of strong Coulomb blockade
$k_BT\ll e^2/C$, this yields\cite{Beenakker91}
\begin{eqnarray}
\label{eq:Gmax}
g_N & = & \frac{4}{\langle \Gamma \rangle}
\sum_{p=N-\frac{L}{2}}^{N+\frac{L}{2}} \frac{ \Gamma_p^l \Gamma_p^r}{
\Gamma_p^l + \Gamma_p^r} F_{\mbox{\scriptsize{eq}}} (\varepsilon_p |
N) \nonumber \\ & & \times [1 - f( \varepsilon_p - \varepsilon_N) ]
\nonumber \\ & & \times f \biglb( -T S(N) + T S(N-1) \bigrb).
\end{eqnarray}
One further remark about the implementation of
Eq.~(\ref{eq:partitionfunction}) is in order. In our numerical
simulations we varied the number of eigenstates $L$ summed over
depending on the temperature range. For $k_BT \le 0.1\Delta$, we found
that as few as 4 were sufficient to achieve good convergence; for
higher temperatures, $k_B T\ge \Delta$, we used as many as 20.

In the experimental setups where plunger gates are used to define both
shape and depth of the confining potential, one expects that
variations in the potential $V_g$ will continuously deform the
dot\cite{Hackenbroich97} This situation is met in lateral quantum
dots, where most of the data were taken
\cite{Chang96,Folk96,Simmel97,Patel97,Patel98} (an important exception
is Ref.~\onlinecite{Simmel99}). Furthermore, as pointed out in the
introduction, as one adds electrons to the dot the mean field also
changes. Thus, to establish theoretically the direct relation between
$V_g$ and the dot electronic mean field it is necessary to know the
experimental set up in detail and perform rather sophisticated
numerical simulations. While such study is desirable, it is not
crucial for our analysis and will not be pursued here. We therefore
simplify the problem parameterizing the ``deformations'' by a generic
variable $X$ on which eigenvalues and eigenfunctions will depend. In
summary, the key point behind this hypothesis is that by adding one
extra electron to the dot the mean electrostatic potential changes
parametrically, on average, by $\delta X$.\cite{Vallejos98} In this
way, the $N$th conductance peak involves a sum over eigenergies and
partial widths $\{\varepsilon_p(X), \Gamma_p^r(X), \Gamma_p^l(X) \}$,
while the next peak in the sequence will be depend on
$\{\varepsilon_p(X+\delta X), \Gamma_p^r(X+\delta X),
\Gamma_p^l(X+\delta X) \}$. It is convenient to measure the
deformations in units of the typical distance between energy level
anti-crossings, namely, the inverse root mean square derivative of the
energy levels with respect to deformations:\cite{Simons93}
\begin{equation}
X_c = \Delta \left[ \left\langle \left( \frac{d\varepsilon_p}{dX}
 \right)^2 \right\rangle - \left\langle \frac{d\varepsilon_p}{dX}
 \right\rangle^2 \right]^{-1/2}.
\end{equation}
(Here $\langle \cdots \rangle$ stands for an average over states $p$
and over the parameter $X$). When the addition of a single extra
electron is enough to scramble strongly energy levels and wave
functions, $\delta x \equiv \delta X/X_c \ge 1$. In this
circumstances, we expect each new conductance peak to signal a
different and independent realization of the effective confining
potential in the dot. Thus, in the regime of strong deformations,
peaks heights should be fairly uncorrelated, in contrast with what
happens as temperature is increased. A systematic study of temperature
and shape deformation combined is carried out below.

 
\section{The effect of parametric variations on thermal averages} 
\label{sec:rmt}
 

In this section we investigate the effect of parametric Hamiltonian
variations on thermal averages. More specifically, we are interested
in answering two questions: (a) Are the correlations between
consecutive peak heights sensitive to shape deformations? (b) What is
the effect of parametric variations on the distribution of peak
heights? We assume that the electronic underlying classical dynamics
is fully chaotic and, therefore, its quantum Hamiltonian can be
modeled by a random matrix. In doing so, we restrict our analysis to
universal aspects; we leave the study of some specific dynamical
models and nonuniversal features to the following sections. Since a
fully analytical treatment of the problem is not available, we
implement the numerical procedure discussed in the
Sec. \ref{sec:thetheory} to evaluate the canonical averages from a
given sequence of energy levels and tunneling rates.

The set of eigenenergies and eigenfunctions of a random parametric
Hamiltonian can be conveniently obtained from a Hermitian matrix of
the form
\begin{equation} 
H(X) = H_1 \cos X + H_2 \sin X, 
\end{equation}
where $H_1$ and $H_2$ are independent $N\times N$ matrices ($N = 500$
in our calculations) belonging to the proper Gaussian ensemble. In our
numerical calculations, the orthogonal ensemble (GOE, $\beta=1$) was
used to model the situation when no magnetic field is present
(time-reversal symmetric systems). For the case with non-zero magnetic
field and broken time-reversal symmetry, we used instead the unitary
ensemble (GUE, $\beta=2$). For each realization of $H_1$ and $H_2$ and
a given value of the parameter $X$, the matrix $H(X)$ was diagonalized
through standard methods. Only the 100 central eigenvalues were kept
after each diagonalization. This was done in order to avoid large
density variations in the spectra and the consequent need for
unfolding, as well as the inclusion of localized eigenstates from the
tails of the band. The resulting eigenfunctions $\{\psi_p(k;X)\}$ were
used to generate the right and left tunneling rates according to the
point contact approximation $\Gamma_p^r (X)\propto |\psi_p(k=1;X)|^2$
and $\Gamma_p^l(X) \propto |\psi_p(k=N;X)|^2$. The energy eigenvalues
$\{\varepsilon_p(X)\}$ were used in the quadrature approximation for
the canonical partition function given by
Eq.~(\ref{eq:partitionfunction}). Thus, for a given realization of
$H_1$ and $H_2$, a certain deformation $\delta X$, and temperature
$T$, we simulated sequences of dimensionless conductance peak heights
as given by Eq.~(\ref{eq:Gmax}).

It is noteworthy that the calculation shown in
Ref.~\onlinecite{Alhassid98}, although similar in formulation to ours,
relied only on the randomness of the tunneling rates and neglected
fluctuations in the eigenenergies by adopting a picket-fence
approximation for the spectrum. Here, instead, we use eigenenergies
fully consistent with the Wigner-Dyson statistics predicted from RMT.
For the unitary case, due to the strong level repulsion the
picket-fence is a reasonable first approximation for the spectrum and
we expect our results to be similar to those of
Ref.~\onlinecite{Alhassid98}. Indeed we observe only small, 10\%
differences in peak height correlation functions (see below). For the
orthogonal case, however, level repulsion is weaker and to obtain
accurate results it is recommended to use the exact spectrum. This
issue becomes very important when calculating
$F_{\mbox{\scriptsize{eq}}}(\varepsilon_p|N)$ for dynamical
systems. In Sec.~\ref{sec:dynamical} we shall resume this discussion.

The results presented here involved data obtained from 50 independent
realizations of the matrices $H_1$ and $H_2$. For each of these
realizations, 150 values of $X$ were used within the interval
$[0,\pi/2]$ to generate peak sequences related to different ``shape''
deformations. Additional statistics was obtained by varying the
``chemical potential'' (i.e., the initial number of electrons in the
dot) and the ``original shape'' (i.e, the initial $X$ point), taking
care to avoid introducing spurious correlations between distinct peak
sequences.

\begin{figure}
\setlength{\unitlength}{1mm}
\begin{picture}(120,150)(0,0)
\put(20,0){\epsfxsize=10cm\epsfbox{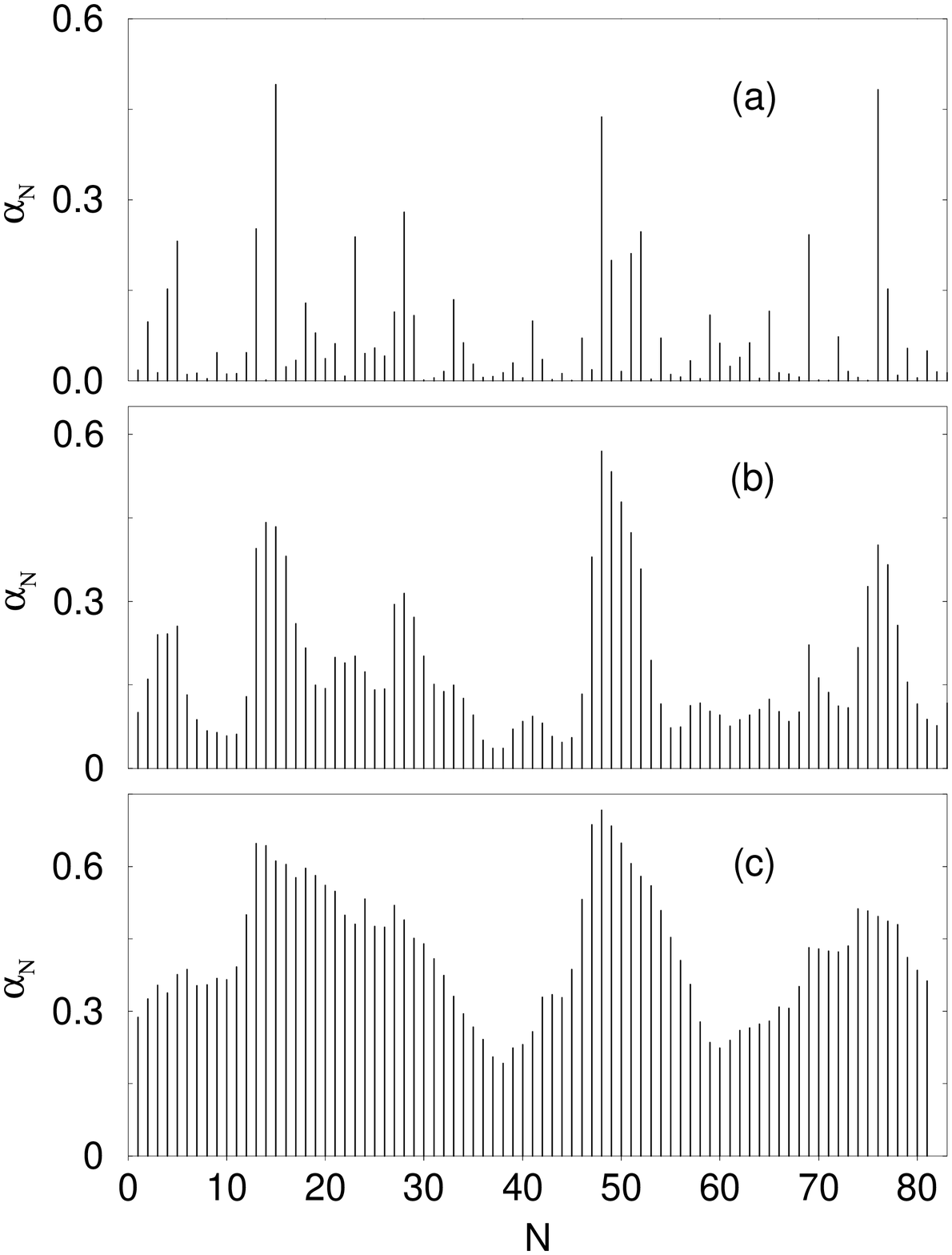}}
\end{picture}
\caption{Sequences of dimensionless conductance peak maxima $\alpha_N$
taken from the GOE simulations for a given set of eigenstates and no
deformation at (a) $k_BT=0.1\Delta$, (b) $k_BT=\Delta$, and (c)
$k_BT=3\Delta$.}
\label{fig:rmtpeaks}
\end{figure}

As an illustration, three typical sequences of conductance peak maxima
are shown in Fig.~\ref{fig:rmtpeaks} for the orthogonal ensemble with
some arbitrary deformation. Notice that, as temperature increases, the
overall modulation becomes more pronounced, marking the existence of
large peak-to-peak correlations. The clustering of peaks with similar
heights is characteristic of the dominance of a single well-connected
eigenstate over a wide energy range.

The most characteristic feature of the thermal average is that the
occurrence of small peaks becomes unlikely, even for temperatures
which are smaller than $\Delta$. This can be observed in the curves
presented in Ref.~\onlinecite{Alhassid98} for the distribution of peak
heights, although the effect is not always quantitatively confirmed by
the experiments. For instance, in the case of preserved time-reversal
symmetry the absence of small peaks should be very pronounced, in
clear contradiction with the currently available experimental
data.\cite{Chang96,Folk96}

\begin{figure}
\setlength{\unitlength}{1mm}
\begin{picture}(80,100)(0,0)
\put(20,0){\epsfxsize=10cm\epsfbox{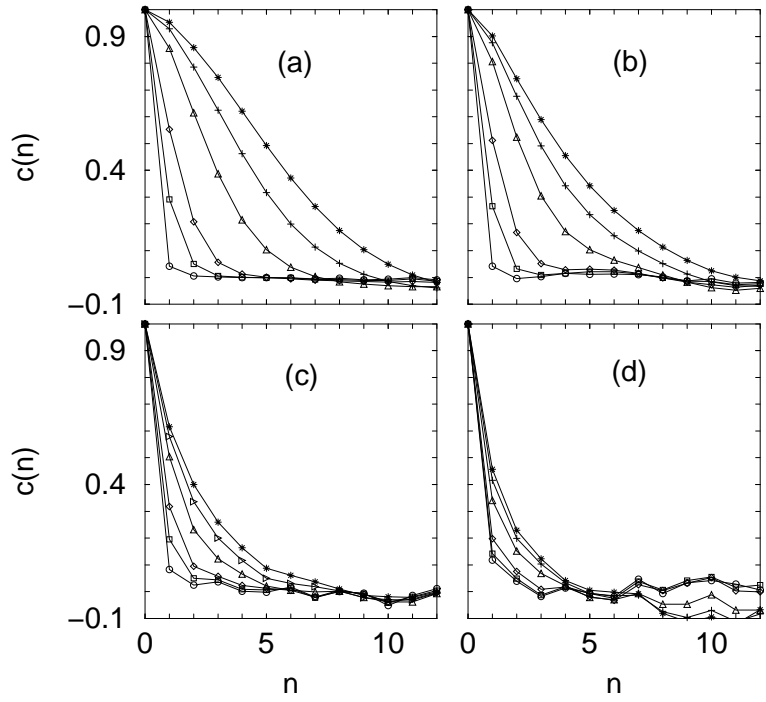}}
\end{picture}
\caption{Peak height correlation function for the GOE at temperatures
$k_B T=0.1 \Delta$ (circles), $0.3 \Delta$, $0.5 \Delta$, $\Delta$,
$1.5\Delta$, $2\Delta$ (stars) for different shape deformations:
(a) $\delta x = 0$, (b) $\delta x = 0.1$, (b) $\delta x = 0.5$, and
(d) $\delta x = 1.0$.}
\label{fig:rmtcorO}
\end{figure}

\begin{figure}
\setlength{\unitlength}{1mm}
\begin{picture}(80,100)(0,0)
\put(30,10){\epsfxsize=9cm\epsfbox{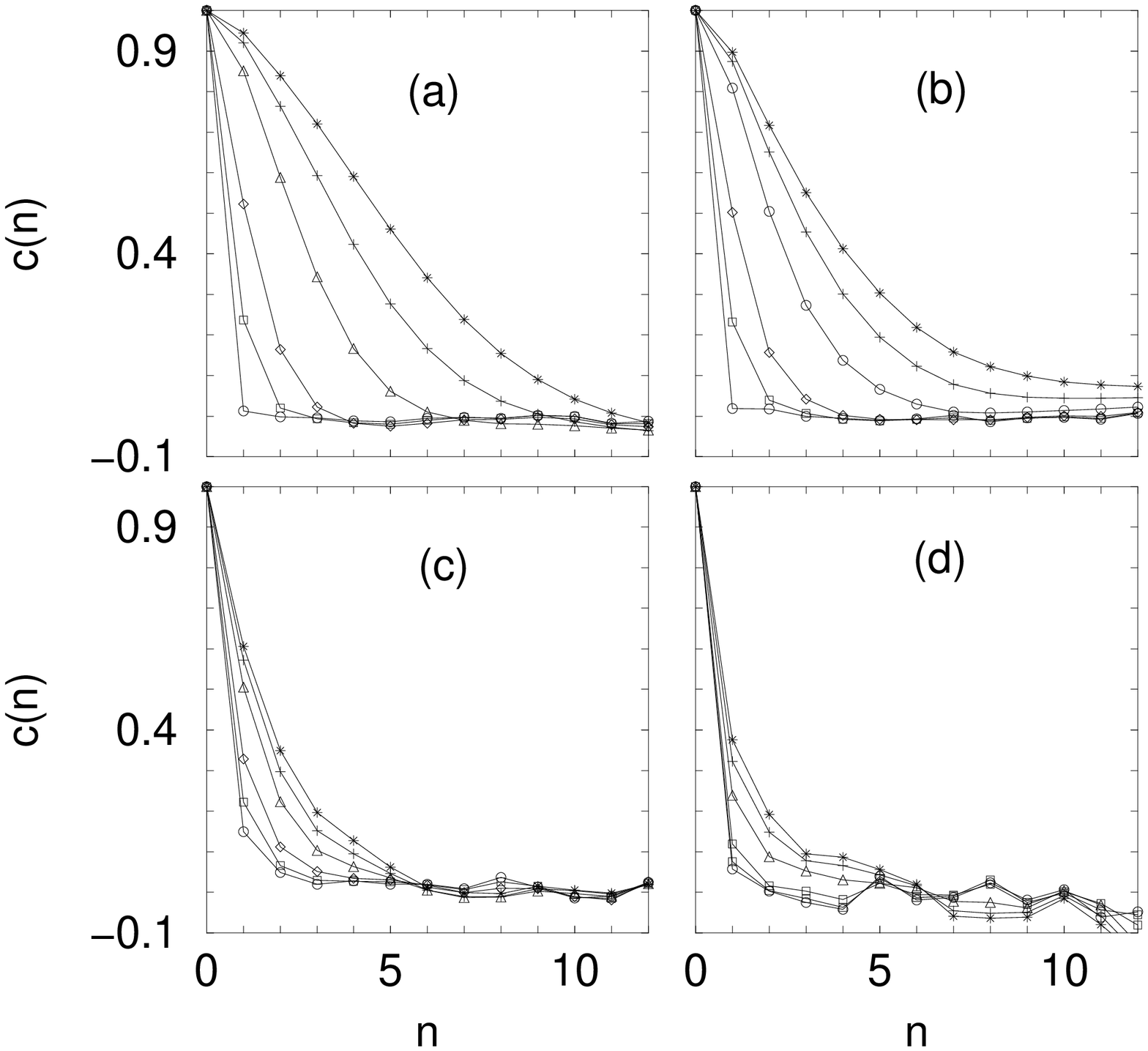}}
\end{picture}
\caption{Same as in \protect{Fig.~\ref{fig:rmtcorO}}, but for the
GUE.}
\label{fig:rmtcorU}
\end{figure}

\begin{figure}
\setlength{\unitlength}{1mm}
\begin{picture}(60,130)(0,0)
\put(40,15){\epsfxsize=6cm\epsfbox{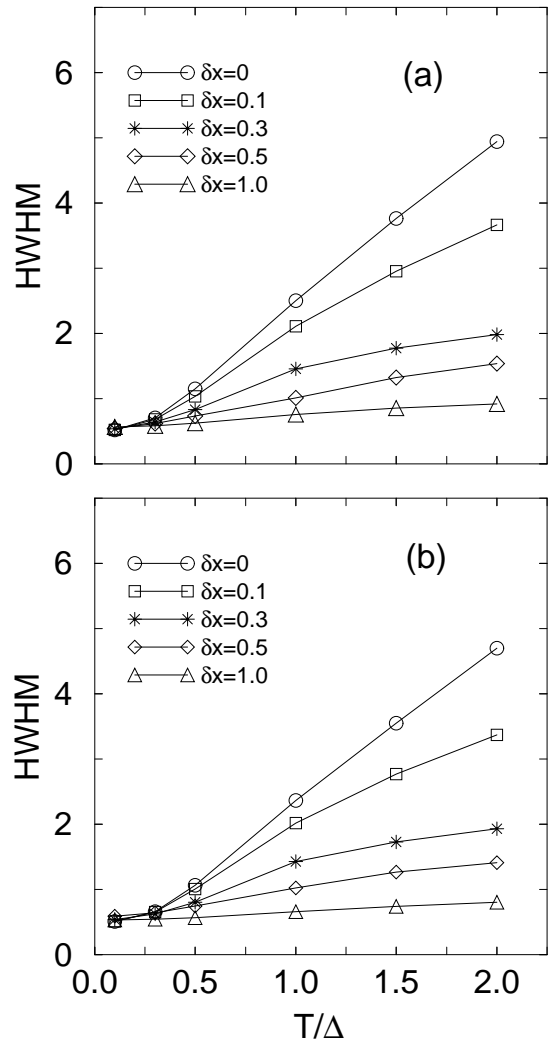}}
\end{picture}
\caption{Correlation length (HWHM) obtained from
Figs.~\ref{fig:rmtcorO} and \ref{fig:rmtcorU} for (a) GOE and (b)
GUE.}
\label{fig:rmtnc}
\end{figure}

To start a quantitative analysis, let us first concentrate on the
correlation of peak heights between neighboring conductance peaks. For
this purpose, we have calculated the correlation function
\begin{equation}
c(n) = \frac{ \left\langle \delta \alpha_{N+n} \delta \alpha_N
\right\rangle } { \left\langle \delta \alpha_N \delta \alpha_N
\right\rangle },
\end{equation}
where $\delta \alpha_N = \alpha_N - \langle\alpha_N \rangle$
represents the deviation from average height of the $N$th peak. The
averages $\langle \cdots \rangle$ are taken over different
realizations of $H_1$ and $H_2$, as well as over $N$. The results for
each Gaussian ensemble at different temperatures and deformations are
shown in Figs.~\ref{fig:rmtcorO} and \ref{fig:rmtcorU}. Notice that
even small parametric changes in the Hamiltonian can rapidly destroy
the correlation of peak heights obtained from thermal averaging. This
effect can be made more quantitative by calculating the half width at
half maximum (HWHM) of the correlator $c(n)$ as a function of
temperature (see Fig.~\ref{fig:rmtnc}).

Small deformations $\delta x$ exclusively due to shape variations are
reported in a recent experiment \cite{Patel98} where, in addition, a
theoretical curve solely based on ``static'' Gaussian ensembles
($\delta x =0$) is used to explain the data. Although the parametric
deformation for this experimental set up is relatively small, one
should not disregard its importance: For $\delta x \approx 0.3$ we
already observe in Fig.~\ref{fig:rmtnc} a much slower increase in
correlation versus temperature than for $\delta x \approx 0$, in
agreement with the data of Ref.~\onlinecite{Patel98}. A fully
consistent, parametric RMT for the peak height fluctuations yields a
{\it smaller} correlation length than what is predicted by the
``static'' Gaussian ensembles. This is a clear indication that the
theoretical curve shown together with the correlation length data in
Ref.~\onlinecite{Patel98} should be taken with
reservation. Nevertheless, the inclusion of parametric deformations
alone does not solve the discrepancy between the RMT predictions and
the experimental data at small temperatures.

\begin{figure}
\setlength{\unitlength}{1mm}
\begin{picture}(60,90)(0,0)
\put(20,0){\epsfxsize=9cm\epsfbox{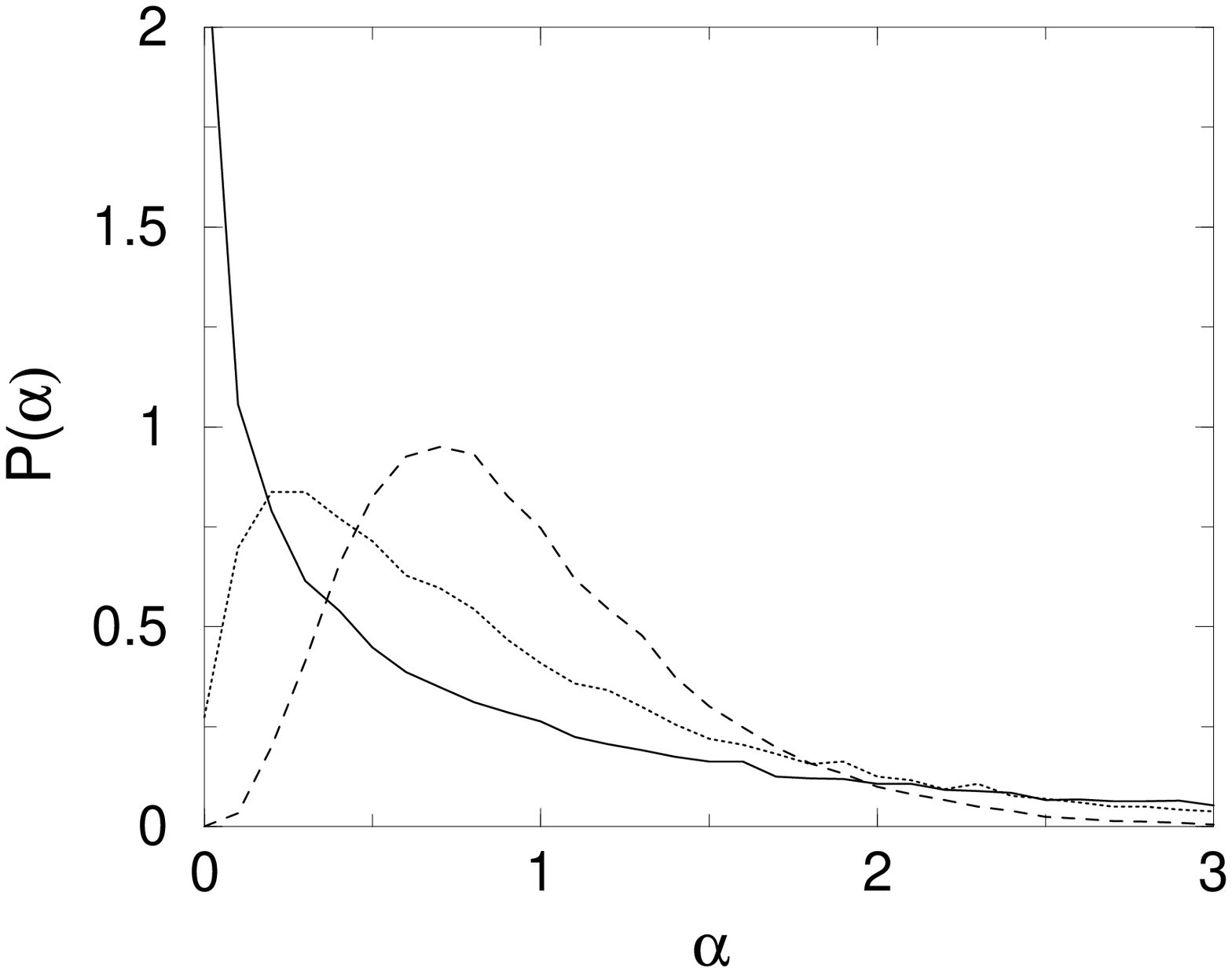}}
\put(68,38){\epsfxsize=4cm\epsfbox{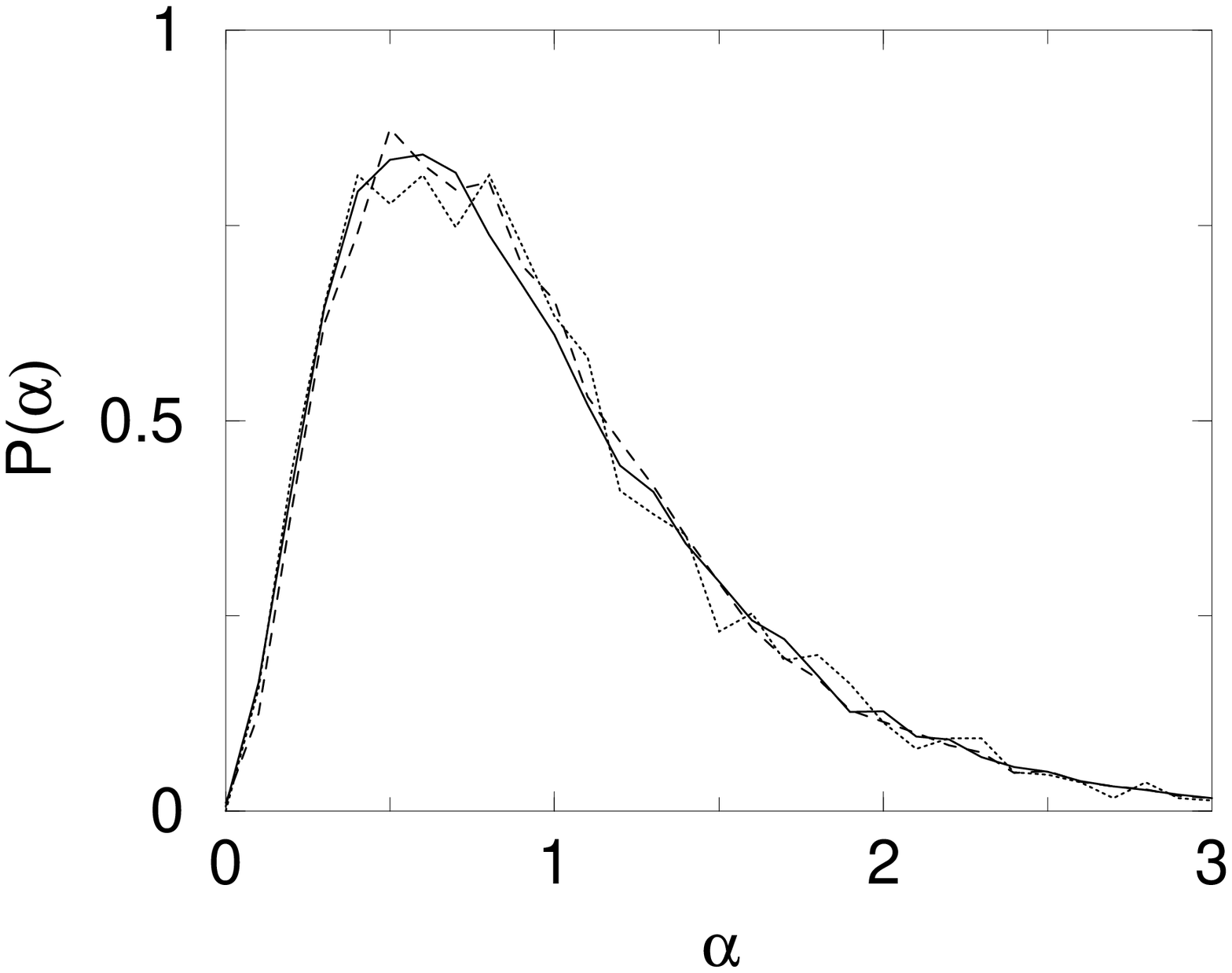}}
\end{picture}
\caption{Distribution of peak heights for the GOE without parametric
distortion at different temperatures: $k_B T=0.1\Delta$ (solid line),
$0.5\Delta$ (dotted line), and $1.5\Delta$ (dashed line). The inset
shows the distribution at $T=0.5\Delta$ obtained for three different
deformations ($\delta x = 0.1$, $0.3$, and $1.0$). The heights are
rescaled to $\langle \alpha \rangle = 1$.}
\label{fig:rmtdistO}
\end{figure}

\begin{figure}
\setlength{\unitlength}{1mm}
\begin{picture}(60,90)(0,0)
\put(20,0){\epsfxsize=9cm\epsfbox{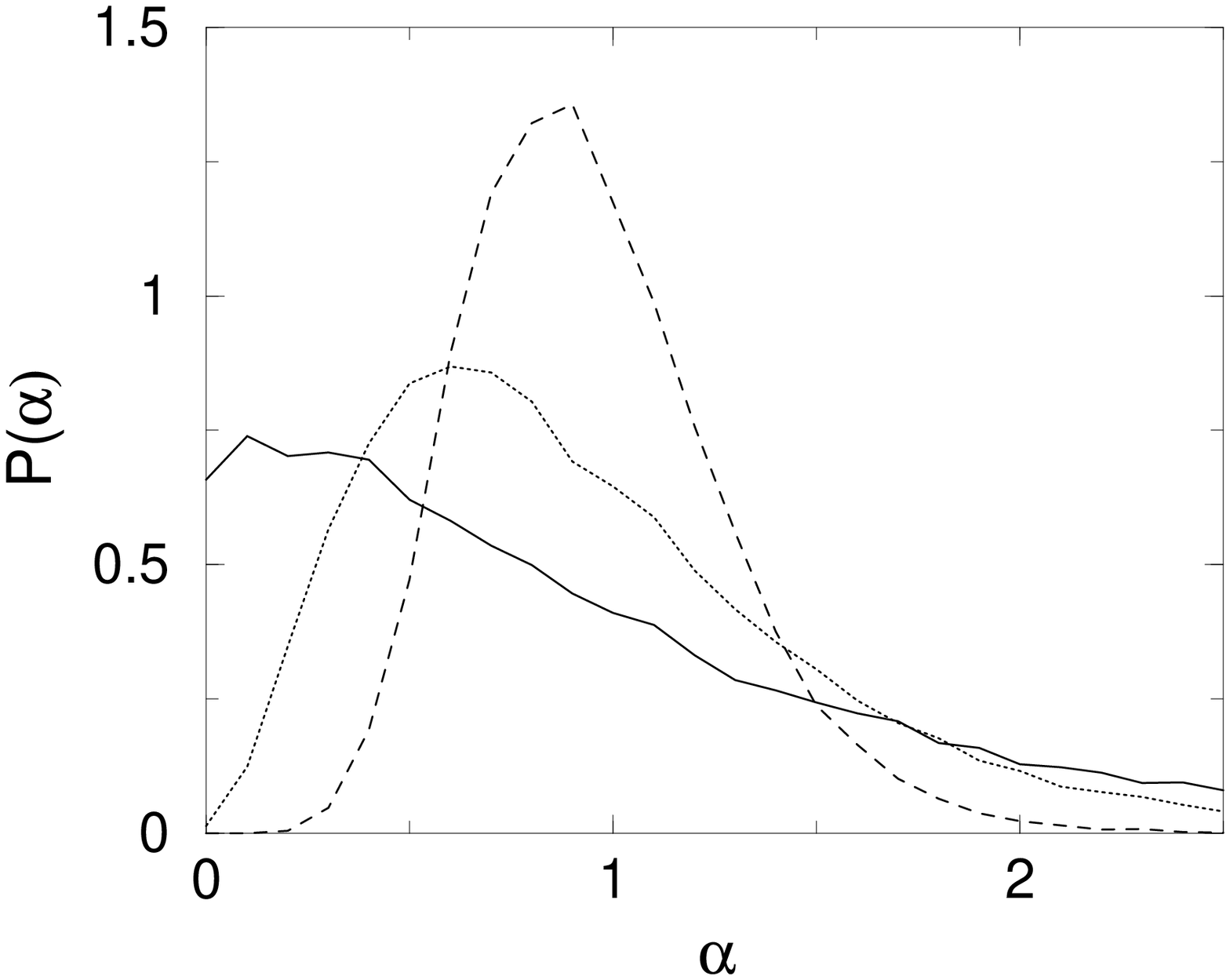}}
\put(68,38){\epsfxsize=4cm\epsfbox{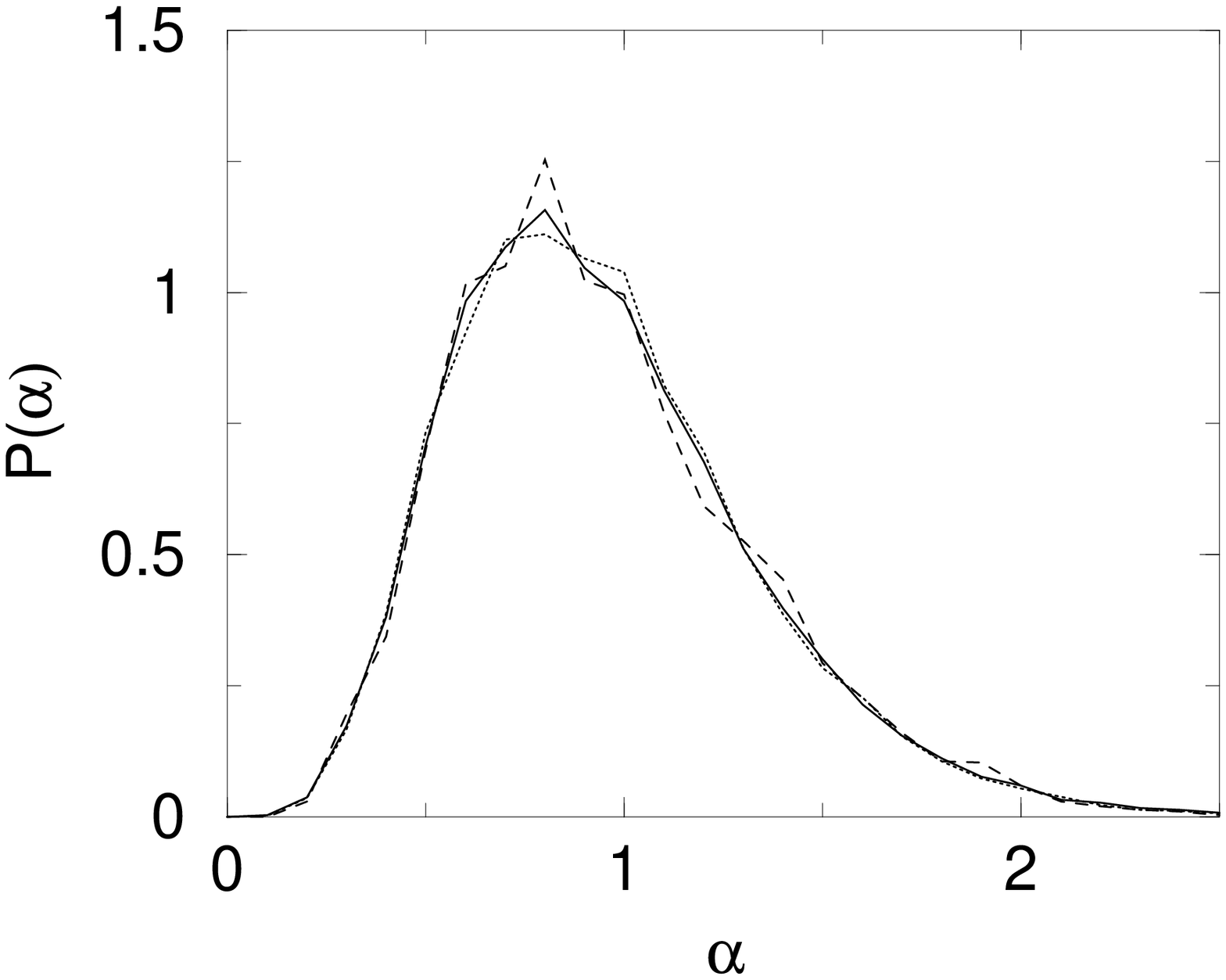}}
\end{picture}
\caption{The same as in \protect{Fig.~\ref{fig:rmtdistO}} but for the
GUE.}
\label{fig:rmtdistU}
\end{figure}

Another important statistical measure of the peak height fluctuation
is its distribution. In Figs.~\ref{fig:rmtdistO} and
\ref{fig:rmtdistU} we show the distributions for both GOE and GUE
cases. The inset illustrates the dependence on deformations for a
given temperature. We observe that, contrary to the correlation
lengths of peak heights, the distributions are not affected by
parametric deformations, but do depend strongly on temperature. For
$k_B T\gg\Delta$ the distribution moves towards a Gaussian shape. It
is important to notice that RMT captures the qualitative aspects of
the temperature dependence observed experimentally,\cite{Patel98} but
fails quantitatively: The data shows a smaller fluctuation in peak
heights at high temperatures ($k_BT> \Delta$) than the theory
predicts. This could be interpreted as an indication that for such
temperatures many-body (or dephasing) effects become
important.\cite{Patel98}

We conclude that a theory based solely on random matrices with a
proper thermal average is not capable of explaining simultaneously the
peak height distribution and correlations. Even before attempting to
improve the theory by including many-body interaction effects, we
shall investigate whether nonuniversal aspects of the underlying
single-particle dynamics also play an important role and lead to
additional peak height correlation mechanisms.

 
\section{Dynamical correlations in chaotic systems} 
\label{sec:dynamical}

 
In this section we investigate how single-particle dynamical
correlations statistically influence the conductance peak heights. We
first examine how universal chaotic wave function correlations
\cite{Berry77,Prigodin95a,Prigodin95b,Srednicki96} manifest on the
conductance autocorrelation function. These correlations go beyond RMT
theory, but are universal in the sense that they arise for any chaotic
quantum state with finite wavelength. We show analytically that at
$T=0$ the effect is small, but is enhanced at higher temperatures. We
also investigate if peak height bunching can be explained by
eigenstates localized in coordinate space. Such localization
corresponds to wave functions with amplitudes concentrated along
certain periodic orbits of the underlying classical Hamiltonian. This
non-universal, spatial structure is known in the quantum chaos
literature as a ``scar"\cite{Heller84} of the classical dynamics on
the wave functions.

We illustrate our investigation by studying numerically a suitable
dynamical model. Namely, we calculate $P(\alpha)$ and $c(n)$ for the
smooth stadium potential as a function of $T$ and deformation. We also
model the coupling between the dot and the leads in two distinct
ways. Most of the results presented below assume point contacts, in
which case $\Gamma^c_p \propto |\psi_p(\bbox{r}_c)|^2$, where
$\psi_p(\bbox{r}_c)$ is the $p$-th dot eigenfunction, evaluated at
some point $\bbox{r}_c$ in the interface region between dot and
lead. This description implies that, close to the dot, the leads
should be narrow and carry a single propagating
channel. Alternatively, we use the standard $R$-matrix formalism to
calculate the decay width for extended leads by taking the overlap
integral over the contact area: $\Gamma^c_p \propto |\int d\bbox{s}
\chi^c(\bbox{r}) \psi_p(\bbox{r}_c)|^2$, where $\chi^c$ is the channel
$c$ wave function. Both models assume that the barrier penetration
factors are small, such that $\langle \Gamma \rangle \ll
\Delta$. Furthermore, it is assumed that the penetration factors vary
smoothly with energy and can thus be incorporated in $\langle \Gamma
\rangle$ without affecting fluctuations.

\subsection{Corrections to the conductance fluctuations due to 
wave function correlations }
\label{sec:weak} 

A recent work by Narimanov and collaborators \cite{Narimanov98}
explored the idea that the short time dynamics can influence peak
height conductance fluctuations. Using the semiclassical theory they
derived a correction to the conductance in the absence of magnetic
field and $k_B T = 0$ which provides a particularly strong effect on
the tails of the autocorrelation function. One of the important
constraints in Ref.~\onlinecite{Narimanov98} is that the results are
obtained for symmetrically placed leads, such that
$\Gamma^r=\Gamma^l$. Our study addresses a more generic situation and
has a different starting point than theirs.

In 1977, Berry \cite{Berry77} suggested that wave functions of
billiards with a chaotic underlying dynamics can be statistically
described by a random superposition of plane waves with a fixed
energy. As a result, he found that chaotic wave functions at energy
$E$ display universal spatial correlations. In two dimensions the
result is
\begin{equation}
\label{eq:wfcor}
C(\bbox{r}_1,\bbox{r}_2;E) = {\cal{A}}^{-1} J_0(k |\bbox{r}_1 -
\bbox{r}_2|),
\end{equation}
where ${\cal{A}}$ is the billiard area, $\hbar k = (2m^*E)^{1/2}$, and
$J_0(x)$ is the ordinary Bessel function. Numerical verifications of
Eq.~(\ref{eq:wfcor}) usually use an arbitrary wave function, fix one
coordinate and average over all directions to obtain $C(\delta r, E)$,
with $\delta r=|\bbox{r}_1-\bbox{r}_2|$. Alternatively, and closely
related to this study, $C(\delta r,E)$ can be obtained by keeping
$\bbox{r}_1$ and $\bbox{r}_2$ fixed and taking the average over wave
functions corresponding to eigenenergies close to $E$ (see for
instance, Ref.~\onlinecite{Alhassid97} and references therein).

Since wave functions are spatially correlated, so are $\Gamma^l$ and
$\Gamma^r$. The joint probability distribution of $\Gamma^r$ and
$\Gamma^l$ has already been obtained in the the orthogonal case
$(\beta=1)$ by the supersymmetric technique.\cite{Prigodin95a}
Srednicki \cite{Srednicki96} later derived the same quantity by
elegantly extending the formalism presented in
Ref.~\onlinecite{Alhassid97}. We can summarize the result in the
following way: Within the point contact model, defining $v^{r(l)}
\equiv \Gamma^{r(l)}/ \langle\Gamma\rangle$, one
finds\cite{Prigodin95a,Srednicki96}
\begin{eqnarray}
\label{eq:pgggoe}
P_{\beta=1}(v^r, v^l) & = & \frac{1}{2\pi (1-f^2)^{1/2} (v^r
       v^l)^{1/2}} \exp \left[- \frac{v^r + v^l}{2(1 - f^2)}\right]
       \nonumber \\ & & \times \cosh \left(\frac{2f\sqrt{v^r v^l}}{1 -
       f^2} \right),
\end{eqnarray}
where $f = {\cal{A}} C(\delta r, E)= J_0(k\delta r)$ in two
dimensions. For $f=0$ one immediately realizes that the standard
product of Porter-Thomas distributions is recovered. Unfortunately, we
did not succeed in computing analytically the conductance peak height
distribution starting from Eq.~(\ref{eq:pgggoe}). However, it is
rather straightforward to express the average conductance as a
function of $f$ in terms of special functions. To leading order in
powers of $f^2$ one finds
\begin{equation}
\langle \alpha (k\delta r)\rangle_{\beta=1} = \frac{1}{4} \left\{1+
\frac{3}{4}[J_0(k\delta r)]^2\right\} + O(f^4).
\end{equation}
For $k\delta r \gg 1$ we recover the average predicted by
Eq.~(\ref{eq:PGOEg}), as we should. In a typical experiment, by adding
electrons and changing $k \delta r$, $\langle \alpha (k \delta
r)\rangle$ becomes a slowly oscillatory function on the energy scale
of peak spacings. These long range oscillations in
$\langle\alpha\rangle$ do not appreciably change the autocorrelation
function $c(n)$ for small $n$, being only pronounced at the tails.
These statement is in agreement with the results shown in
Ref.~\onlinecite{Narimanov98}. By identifying $k = k_F$ (the Fermi
wave number) and taking a typical experimental value for $k_F \delta
r$, the amplitude of the modulation will be of the order 1\% of the
standard RMT value for $\langle \alpha \rangle$. This means that the
maximum anticorrelation in $c(n)$, at $n =
n_{\scriptsize{\mbox{anti}}}$, is also about 1\% of $c(0)$.

A similar calculation can be made for the case of unitary symmetry
$(\beta =2)$. Also in this case the joint probability distribution has
already been derived \cite{Prigodin95b,Srednicki96} and reads
\begin{equation}
\label{eq:pgamgamgue}
P_{\beta=2}(v^r, v^l) = \frac{1}{1 - f^2} \exp \left(- \frac{v^l +
  v^r}{1 - f^2}\right) I_0 \left(\frac{2f\sqrt{v^r v^l}}{1 - f^2}
  \right),
\end{equation}
where $I_0(x)$ is the modified Bessel function. Here, again, it is
difficult to obtain the conductance peak height distribution
$P(\alpha)$ in a closed form. The averaged $\langle \alpha\rangle$ as
a function of $f$, on the other hand, can be found without much
effort. The result in ascending powers of $f^2$ reads
\begin{equation}
\langle \alpha (k\delta r)\rangle_{\beta=2} = \frac{1}{3}
\left\{1 +
\frac{2}{5}[J_0(k\delta r)]^2 \right\}+ O(f^4),
\end{equation}
which is an even smaller correction than for the orthogonal case.

In distinction to the findings of Ref.~\onlinecite{Narimanov98}, such
spatial correlation effects will be hardly observable at zero
temperature. We attribute the difference between our results and
theirs mainly to their stringent constraint of having an exact
symmetry in the wave functions near the tunneling region.
Nonetheless, we anticipate that the long-range autocorrelation
oscillations become more pronounced with increasing temperature, as we
show in our numerical calculations. For future comparison with
experiments, it is important to notice that the presence of an
obstacle between $\bbox{r}_1$ and $\bbox{r}_2$ just requires the
replacement of $\delta r$ by the length of the shortest classical path
connecting $\bbox{r}_1$ to $\bbox{r}_2$.\cite{Hortikar98} Before
concluding, an additional comment is in order. Until the present,
Coulomb blockade experiments have offered very few fingerprints of the
underlying classical electronic dynamics. An experimental check of the
effect proposed in Ref.~\onlinecite{Narimanov98}, exploring the
symmetries of the wave function, would offer a clear indication of the
validity range of the single-particle approximation and the
semiclassical analysis.

 
\subsection{Non-universal dynamical correlations due to scars} 
\label{strong} 
 
Strong wave function correlations due to classical structures in phase
(and coordinate) space, such as scars,\cite{Heller84} can also lead to
a non-universal behavior in the peak height
fluctuations.\cite{Stopa97} The analysis presented in this section is
motivated by a recent work of Hackenbroich and collaborators
\cite{Hackenbroich97} which explored the idea of shape deformation to
explain correlations between heights of neighboring peaks. These
authors proposed a correlation mechanism tailored for integrable
systems. In what follows, we recast and reinterpret their mechanism
for chaotic systems with strong scarring. Our results for this case
are qualitatively distinct from those of
Ref.~\onlinecite{Hackenbroich97}.

For an integrable system it is well known that the structure of a
given wave function does not change upon a parametric variation of the
Hamiltonian. The reason is quite simple. The wave functions of
integrable systems are strongly concentrated around classical tori,
and the latter are only smoothly distorted by a parametric variation
of the Hamiltonian. This behavior is generic, provided the system
stays integrable upon the parametric variation. However, as the
parameter $X$ is varied, levels cross each other. Since they
correspond to different sets of quantum numbers, the typical crossing
distance is much smaller than the distance necessary to appreciably
change the wave functions. Let us assume that in certain system there
are levels very weakly influenced by changes in $X$. We call them
``horizontal" levels $h$. (Any $h$ level certainly has a quite marked
wave function structure with respect to the system geometry.)
``Horizontal" levels may cross generic levels, as the parameter $X$ is
varied. Let us assume now that the conductance peak of a dot
containing $N$ electrons is dominated by a certain $h$ level for some
value of $X$. If the typical $\delta X$ caused by the addition of one
extra electron to the dot is of the order of the average parametric
distance between crossings, the $h$ level will always stay very close
to the Fermi energy. In this way, the sequence of conductance peaks
are dominated by the very same ``horizontal" level and are thus
expected to show correlations in heights. This mechanism
\cite{Hackenbroich97} is robust as long as the system dynamics is
close to integrable. The restrictive condition of having exactly one
anticrossing for $\delta X$ can be relaxed if temperature is
included.\cite{Baltin98}

In the case of chaotic systems, wave functions typically decorrelate
after one level crossing.\cite{Mucciolo95} Deviations of such
universal behavior indicate the presence of scarred eigenstates. For
such wave functions the mechanism described in the previous paragraph
is applicable. However, one should be aware that if peak height
correlations are explained by a non-universal feature they are not
ubiquitous.

In what follows we shall exemplify such mechanism and investigate how
often these situation occurs. Moreover, by changing shape, position of
the contacts, and temperature, we nearly exhaust all sources of peak
height correlations within the single-particle scenario. The results
are compared with those of RMT and the available experimental data.

\subsection{Dynamical model: the smooth stadium potential}
\label{smoothstadium}

In order to investigate both aforementioned mechanisms, we studied the
two-dimensional model Hamiltonian $H=\bbox{p}^2/2 + V(\bbox{r})$ (we
take the electron mass $m^\ast=1$), with the potential given by
\cite{Ozorio}
\begin{equation} 
V(\bbox{r})=\left\{ \begin{array}{cc} y^{2n} & x \le a \\ \Big[(x-a)^2
+ y^2\Big]^n & x \ge a \\
 \infty & x<0\ \mbox{or}\ y<0 
\end{array} \right. .
\end{equation} 
The exponent $n$ sets the steepness of the confining potential. The
model is flexible enough to allow for different classical
behaviors. For instance, if we take $n=1$ the system is integrable. As
we increase the value of $n$ our model Hamiltonian becomes very
similar to the well-known stadium billiard, one of the paradigms of
classical chaotic systems. For $a=1$ the equipotential $V(x,y)=1$
corresponds to the border of the hard walled stadium billiard with
unit radius and unit rectangular length.

\begin{figure}
\setlength{\unitlength}{1mm}
\begin{picture}(80,80)(0,0)
\put(20,-20){\epsfxsize=10cm\epsfbox{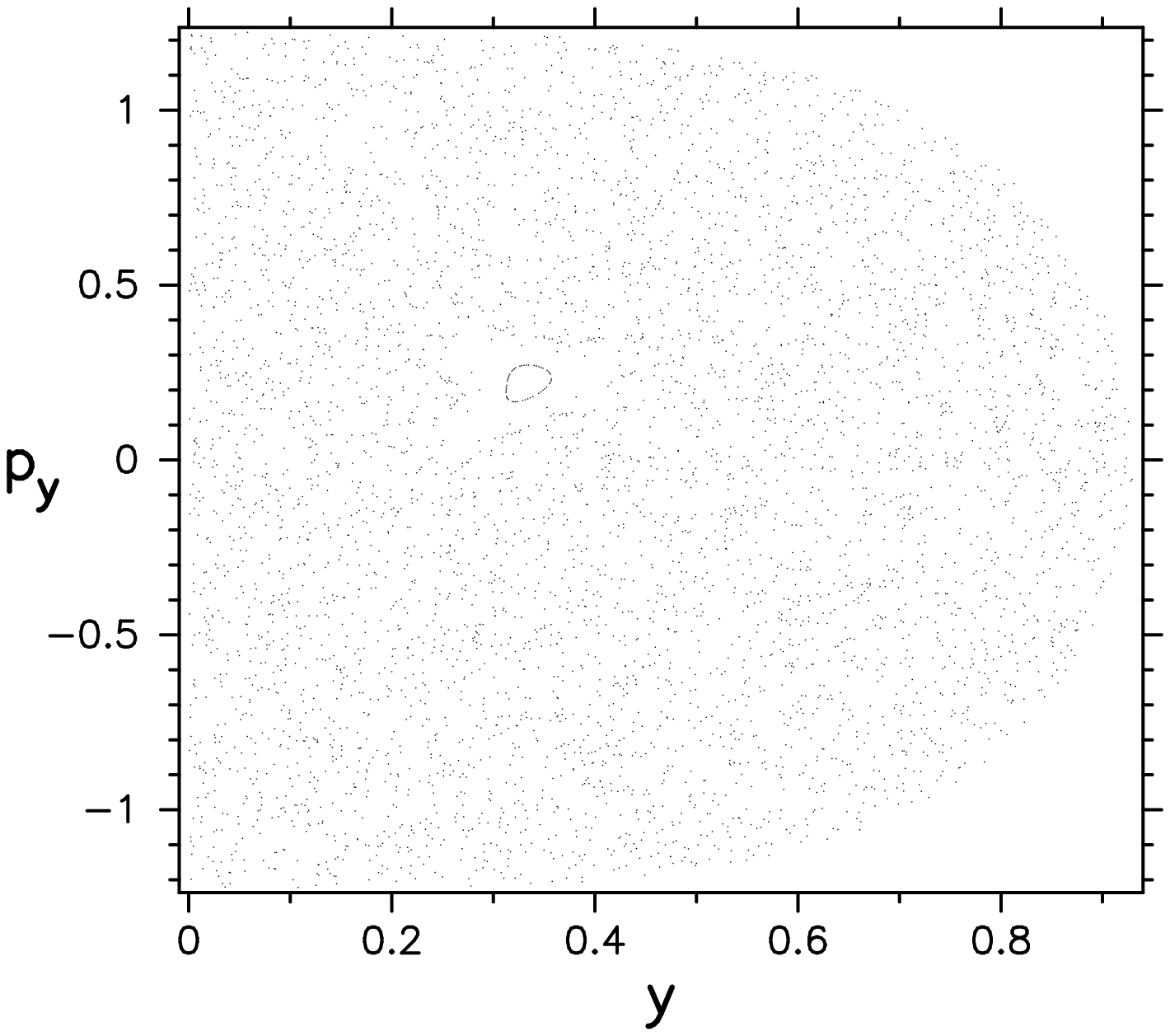}}
\end{picture}
\caption{Poincar\'e surface of section for the smooth stadium billiard
with $n=2$, $a=1$, and $E=0.75$. The section is taken at $x = a$.}
\label{fig:poincare}
\end{figure}

We model changes in the single-particle potential by varying $a$, the
length of the rectangular part of the well. The classical motion is
chaotic if $a$ is varied in the interval range $1.0 \le a \le 1.25$,
provided we choose $n=2$ and stay in the energy window around
$E=1$. Indeed, for this energy and parameter values the classical
phase space is mostly chaotic, as can be seen in
Fig.~\ref{fig:poincare}, though small remnants of integrability still
exist.
 
\begin{figure}
\setlength{\unitlength}{1mm}
\begin{picture}(120,100)(0,0)
\put(20,-30){\epsfxsize=12cm\epsfbox{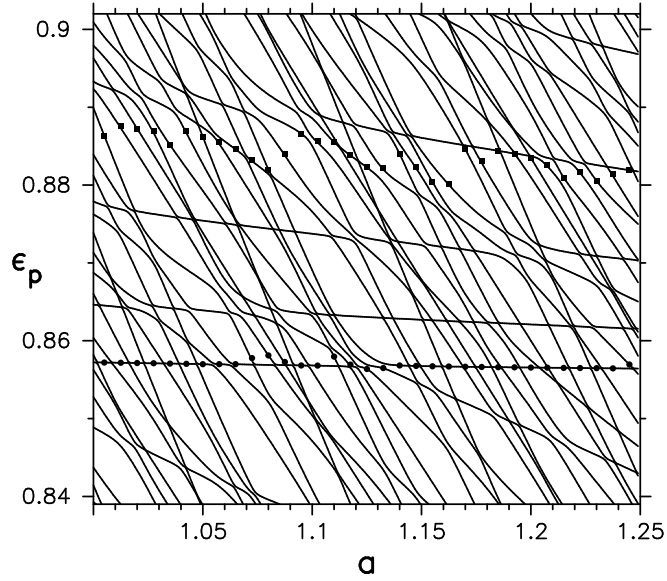}}
\end{picture}
\caption{Energy levels as a function of the length $a$ for the smooth
stadium with $n=2$ and no magnetic field. The two nearly horizontal
sequences of dots correspond to tunneling states at low temperatures
regime and for $\delta x = 1$. The lower energy sequence follows a
scarred pattern, while the higher one does not.}
\label{fig:spaguetti}
\end{figure}

Finally, to assure that the single-particle levels we analyze
correspond to a classically chaotic motion, we fix $\hbar$ in order to
have a window of energy around $E \le 1$ containing typically 100
eigenstates of our Hamiltonian. The quantum eigenstates are obtained
by numerical diagonalization. The problem is solved by using as basis
states the eigenfunctions of a sufficiently large rectangular
box. When the index $n$ is an integer number, the calculation of the
matrix elements is straightforward. The secular matrices are taken
large enough such as to guarantee the convergence of all eigenvalues
we analyze within 1\% of the mean level spacing. This gives us
confidence that the computed wave functions have converged. After
ordering the eigenvalues in ascending energies, we consider only the
eigenstates between the 200th and the 300th levels. A representative
region of the spectrum of the smooth stadium with $n=2$, in the
absence of an external magnetic field, is shown in
Fig.~\ref{fig:spaguetti}.

Quantum calculations including a magnetic field $B$ perpendicular to
the stadium are also of simple implementation in this model. By the
usual minimum ansatz $\bbox{p} \rightarrow \bbox{p} + e
\bbox{A}(\bbox{ r})/c$ with $\bbox{B} = \nabla \times \bbox{A}$ we
obtain a new Hamiltonian that can be diagonalized following the same
numerical procedure. We found useful to work in the gauge where
$\bbox{A} = B/2 [-c_x(y- y_0), c_y(x - x_0)]$ with $c_x+c_y=2$. Of
course, the converged results do not depend on a particular choice of
the gauge. However, the numerics can be made more efficient by
choosing $(x_0, y_0)$ close to the geometrical center of the stadium
and $c_xL_x = c_yL_y$, where $L_x$ and $L_y$ are the lengths of the
classically allowed region of the billiard at $E=1$. With this choice,
we obtain the largest ratio of number of converged eigenvalues to size
of the secular matrix. We followed Ref.~\onlinecite{Pluhar95} to
estimate the magnitude of the time-reversal breaking magnetic flux
$\phi_c$ threading the stadium. At the Fermi energy we obtain
$\phi_c/\phi_0 \approx 2$, where $\phi_0$ is the unit flux
quantum. When comparing the results of the dynamical model against the
RMT unitary case, we fix the magnetic field to be $B {\cal{A}} =
\phi_c$, where ${\cal{A}}$ is the classical allowed area for an
electron with energy $\varepsilon_F$.

\begin{figure}
\setlength{\unitlength}{1mm}
\begin{picture}(100,100)(0,0)
\put(0,-70){\epsfxsize=14cm\epsfbox{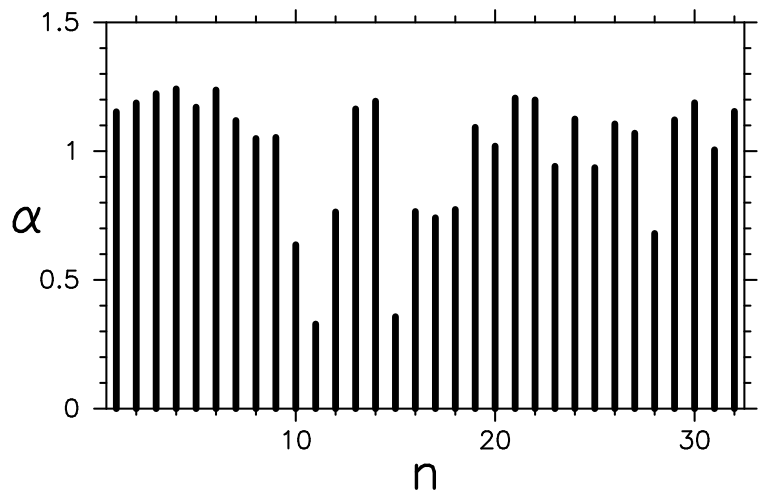}}
\end{picture}
\caption{Sequence of peak heights for ``bouncing ball'' wave functions
at $k_BT = 0.25\Delta$.}
\label{fig:bouncingball}
\end{figure}

Although the system is chaotic, there are many very narrow avoided
crossings, which are characterized by a gap $\epsilon \ll
\Delta$. These crossings occur for only a few eigenstates, indicating
that the latter are very weakly coupled to others eigenstates. This
non-universal feature is well-known in quantum chaos and is a
fingerprint of scarred wave functions. Scarred eigenstates have the
property of concentrating wave function amplitudes along certain
classical periodic orbits. Upon a parametric variation and after a
level crossing, scarred eigenstates preserve their strong identity. In
view of Fig.~\ref{fig:spaguetti}, we can apply the mechanism proposed
for integrable systems in Ref.~\onlinecite{Hackenbroich97} to explain
strong peak-to-peak correlations. We identify the $h$ levels with
scarred ``horizontal" states. The states that contribute to the
conductance in the case where $\delta X = X_c$ and $k_BT\ll \Delta$
are indicated in Fig.~\ref{fig:spaguetti} by dots. The two sequences
shown correspond to different Fermi energies. The wave functions
related to the lower energy sequence are concentrated along the family
of periodic orbits that bounce from $y=0$ to $y=1$ with $v_x=0$ (the
so called bouncing ball modes). The corresponding conductance peak
heights show very large peak-to-peak correlations when the leads are
placed such as to intersect these bouncing trajectories. The result is
shown in Fig.~\ref{fig:bouncingball}. Notice that for small
temperatures the mechanism holds even if the scarred state does not
correspond to the Fermi level, but is its neighbor. This happens
because scarred levels show very weak level repulsion and very often
contribute significantly to the canonical occupation factor at $k_B
T\ll \Delta$. (Differences in $c(n)$ due to spectral fluctuations were
already discussed in \ref{sec:rmt}).

The proposed mechanism holds for chaotic systems if three main
conditions are met. First, the tunneling state at the Fermi energy has
to be localized in coordinate space. Second, $\delta X$ has to be
close to $X_c$ to keep this state close to $\varepsilon_F$.  Third,
the leads have to be in the region where the wave function has an
enhanced amplitude. In practice, these conditions are obviously
difficult to be satisfied simultaneously. However, one could imagine
that a given sequence can have few correlated peaks dominated by a
particular scarred state. When this state drifts away from the Fermi
level it will eventually be replaced by another localized state. In
this way a peak height sequence would have several regions where peaks
are strongly correlated. This is the kind of effect that we actually
tried to identify. Although we observed sequences like that in the
numerical simulations, the proposed mechanism is not generically
robust. When one averages over many sequences by varying the positions
of the contacts inside the billiard, the Fermi level, as well as
$\delta X_c$. The autocorrelation functions $c(n)$ dependence on
$k_BT/\Delta$ does not differ much from the RMT result. In other
words, sequences like the top one in Fig.~\ref{fig:spaguetti} are
statistically dominant.

Summarizing: chaotic systems like the stadium can, in some cases,
display strong peak-to-peak correlations. Such cases, however, are
statistically rare and in our opinion this mechanism is not robust
enough to explain the experimental data.\cite{Folk96,Patel98}

\begin{figure}
\setlength{\unitlength}{1mm}
\begin{picture}(100,90)(0,0)
\put(20,-30){\epsfxsize=12cm\epsfbox{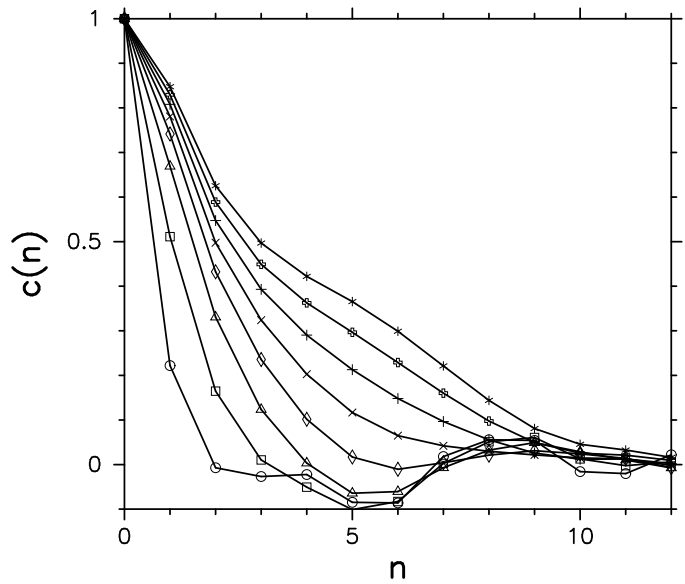}}
\end{picture}
\caption{Peak height autocorrelation function $c(n)$ for the smooth
stadium ($n=2$, no magnetic field) obtained from 60 sequences of 100
peaks each, going from $k_BT = 0.25\Delta$ (circles) to $2\Delta$
(stars) in increments of $0.25\Delta$.}
\label{fig:longcorr}
\end{figure}

Let us now turn to the study of long range correlations. For this
purpose we evaluated $c(n)$ for the smooth stadium without magnetic
field. Deformations which cause energy levels to change by $\Delta$
are very small at the classical level and do not influence our
results. We therefore only consider the case of $\delta X=0$. The data
set consists of 60 sequences of 100 peaks each. The peaks in each
sequence are normalized to unit average height. This is done to avoid
the systematic errors due to the decrease of the mean conductance from
one sequence to another (since each sequence corresponds to a
different length $a$). A correlator is calculated for each sequence
and an average over all sequences is taken. For $k_BT=0$ no
undershooting is observed in $c(n)$, as predicted in
Sec.~\ref{sec:weak}. The results for $k_BT=0.5\Delta$ are shown in
Fig.~\ref{fig:longcorr}. We do not know how to estimate the magnitude
of the oscillations in $c(n)$ after the temperature average, since $f$
enters in the expressions in a convoluted manner. In
Fig. \ref{fig:longcorr} we observe much larger oscillations than
predicted by the theory at $k_BT=0$. We can check if the enhanced
oscillations are related to the mechanism proposed in \ref{sec:weak},
by verifying if $n_{\mbox{\scriptsize anti}}$ (the position of the
anticorrelation maximum) is consistent with our theory. Our data were
obtained by sweeping $k_F$ and keeping the lead positions fixed. For
spinless electrons in a billiard, $k_F \approx (4\pi
N/{\cal{A}})^{1/2}$. To add one electron to the dot, we have to move
up in energy by $\Delta$ in our model. This corresponds to change
$k_F$ by $2\pi/{\cal{A}}k_F$. Using the parameters of our
calculations, we then estimate the position of the maximal
anticorrelation to be $n_{\mbox{\scriptsize anti}}\approx 8$, which
has to be contrasted with Fig.~\ref{fig:longcorr}. A similar exercise
was done in Ref.~\onlinecite{Narimanov98} using realistic values to
estimate $n_{\mbox{\scriptsize anti}}$ from the experimental data,
obtaining values in reasonable agreement with
Ref.~\onlinecite{Patel98}. In our work, anticorrelations in $c(n)$
cannot be ruled out for small values of $k_BT/\Delta$; nevertheless,
they have to increase at higher temperatures and this is observed in
our numerical simulations. We postpone the presentation of this result
to the next section, where we also discuss the possible statistical
problems inherent to such study.


\section{Finite data set and the peak height autocorrelation
function}

\label{sec:finitedataset}


The proper statistical treatment of experimental data is of great
importance in any fluctuation analysis. In practical situations the
number of Coulomb blockade peaks in a sequence goes from 50 to 100.
Therefore, the problem of biases and errors due to finite sample size
is one of the major difficulties in calculating peak height
autocorrelation functions. If a reliable estimate of the effective
number of independent points $N_{\scriptsize{\mbox{eff}}}$ can be
made, the problem of making corrections and estimating errors is
partially solved.

Considerable effort was devoted to this type of problem in the context
of compound nucleus reactions (see, for instance,
Ref.~\onlinecite{Ericson66} and references therein). In this case the
observable of interest is the nuclear cross section, whose
autocorrelation function is known to be Lorentzian. Let us translate
this earlier estimate for the effective number of independent peaks
into the language of Coulomb blockade. For this purpose, we assume
momentarily that the peak height autocorrelation function has a
Lorentzian shape. Doing so and introducing $r\equiv n/n_c$, we can
write the variance of the average conductance peak height as
\begin{equation}
  \mbox{var} \left( \overline{\alpha} \right) = \mbox{var} \left(
    \alpha \right) \left[ \frac{2\tan^{-1}r}{r} -
    \frac{\ln(1+r^2)}{r^2} \right],
\end{equation}
where $\mbox{var}(\alpha)$ is the variance of the conductance peak
height (either the ``true'' variance or, for an estimate, the ensemble
variance), whereas $\overline{\alpha}$ is the average over a sequence
of $n$ peaks with a correlation length $n_c$. It is known that if an
average of $N_{\scriptsize{\mbox{eff}}}$ independent data points is
computed, the variance of this average is given by
\begin{equation}
\mbox{var}( \overline{\alpha} ) = \frac{\mbox{var}(
\alpha)} {N_{\scriptsize{\mbox{eff}}}}.
\end{equation}
By comparing the two last equations, we can express
$N_{\scriptsize{\mbox{eff}}}$ in terms of $n/n_c$. For values of
$n/n_c\gg 1$, $N_{\scriptsize{\mbox{eff}}}$ is very well approximated
by
\begin{equation}
N_{\scriptsize{\mbox{eff}}} \approx \frac{1}{\pi} \frac{n}{n_c},
\end{equation}
which tells us that an effective independent point is found after an
interval of $\pi n_c$ peaks. The statistical fluctuations are, as
usual, given by $1/\sqrt{N_{\scriptsize{\mbox{eff}}}}$. We expect our
estimate of the number of effectively independent points to change
little when a different functional form for the autocorrelation
function is considered. The factor $\pi$ will be replaced by another
numerical factor of the same order of magnitude.

\begin{figure}
\setlength{\unitlength}{1mm}
\begin{picture}(100,160)(0,0)
\put(10,45){\epsfxsize=12cm\epsfbox{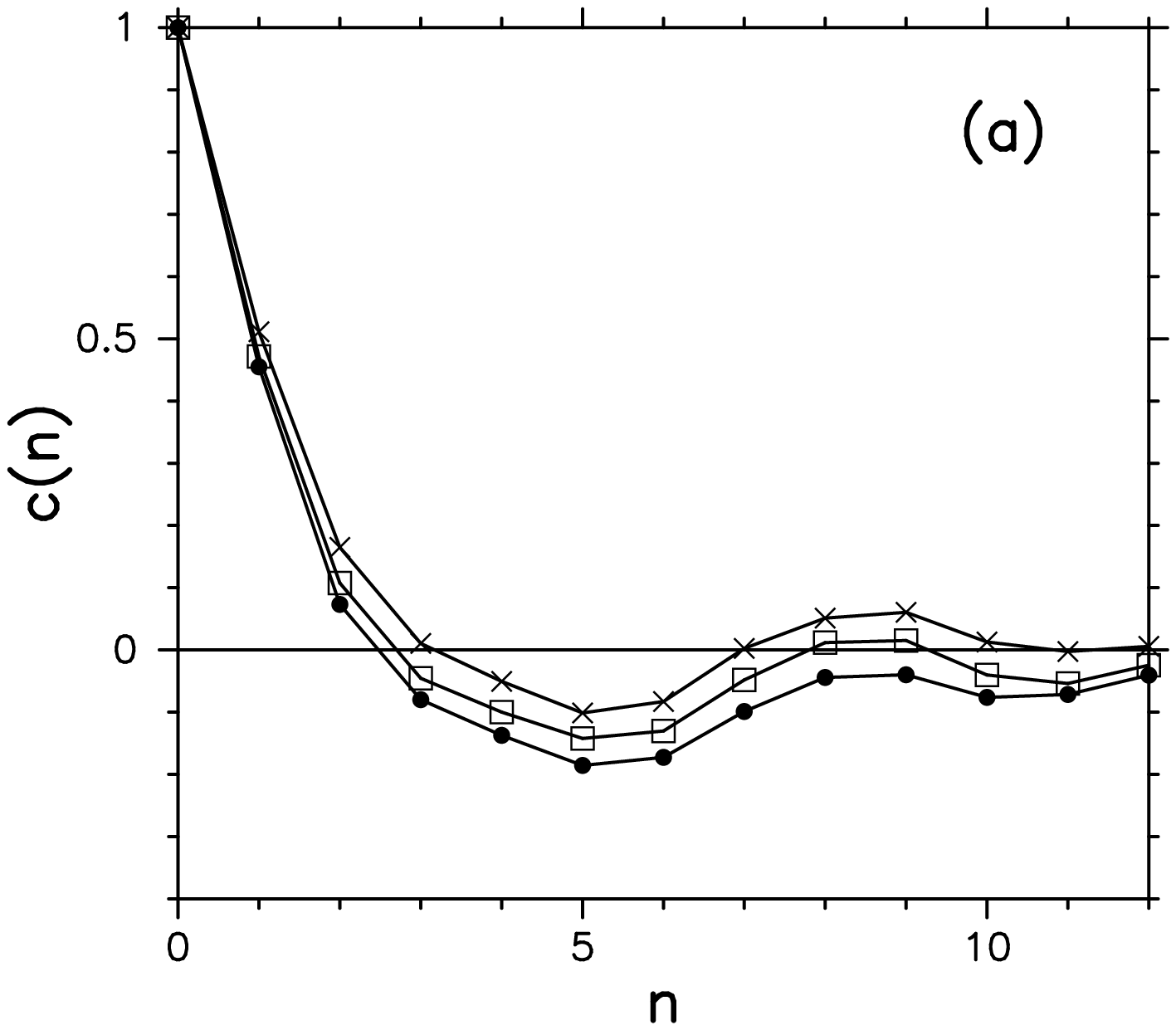}}
\put(10,-35){\epsfxsize=12cm\epsfbox{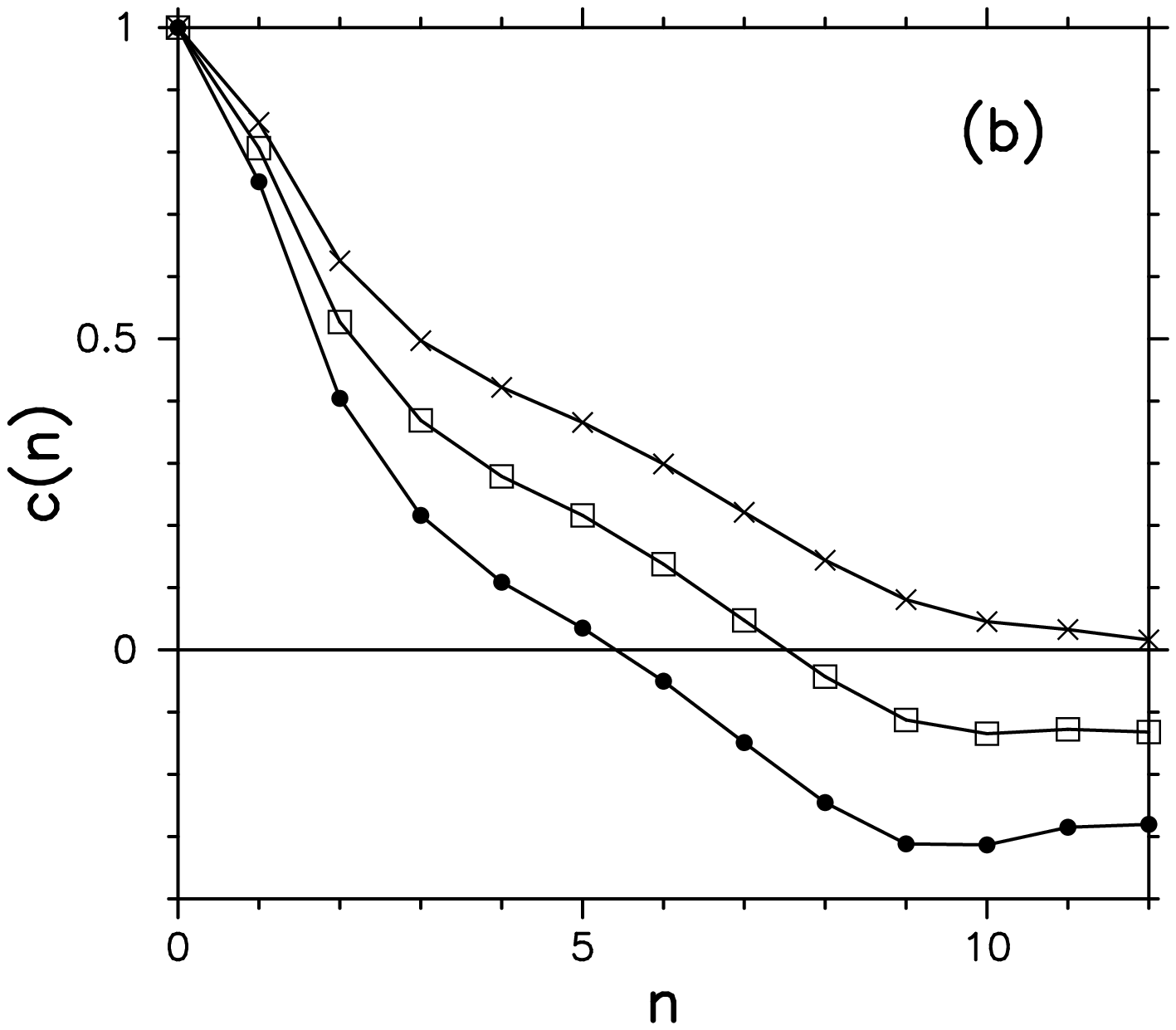}}
\end{picture}
\caption{Peak height autocorrelation function $c(n)$ for the smooth
stadium obtained from different sets of sequences: 60 sequences of 100
peaks (crosses), 120 of 50 (squares), and 240 sequences of 25 peaks
(circles). (a) $k_BT = 0.5\Delta$; (b) $k_BT = 2\Delta$.}
\label{fig:appa2}
\end{figure}

We illustrate this discussion by considering peak height
autocorrelation functions obtained from sequences of different lengths
for the smooth stadium without magnetic field or deformation for $k_BT
= 0.5\Delta$ and $2\Delta$. As temperature increases, so does $n_c$;
statistical fluctuations become large even for samples with many
peaks. This is illustrated in Fig.~\ref{fig:appa2}. At $k_B T =
2\Delta$, $n_c = 5$ and each sequence of 25 peaks has
$N_{\scriptsize{\mbox{eff}}} = 1$ and fluctuations prevail at $n\gg
1$. Similar results are obtained when a magnetic field is included,
but are not shown here.

Notice that the correlation length is more sensitive to small samples
at high temperatures. For instance, at $k_B T = 2\Delta$, $n_c$
decreases appreciably as we consider shorter sequences. However, at
$k_B T = 0.5\Delta$ the correlation length (defined as the HWHM) is
stable. As a consequence, the curve $n_c(T)$ obtained from short
sequences tends to saturate at high temperatures. This is similar to
what is seen in Ref.~\onlinecite{Patel98} for the dot with
$\Delta=38\mu eV$ (Dot 2), the smallest one among the three dots
tested. Our conclusion is that it is rather difficult to affirm
unambiguously that the saturation of $n_c$ with temperature is a
manifestation of decoherence. It may well be the effect of small
statistics.


\section{Are conductance fluctuations a fingerprint of classical
         chaos?}
\label{sec:integravel}


Any systematic search in the literature for single-particle models
designed to explain the fluctuations of Coulomb blockade peak heights
will reveal a strong bias towards the assumption of chaos. Very little
attention has been devoted to integrable models, with, to the best of
our knowledge, one exception.\cite{Bruus94} This is quite
understandable since it is believed that, due to small irregularities
in the confining potential or the presence of weak impurities, the
electronic dynamics tends to be chaotic. In fact, it is particularly
difficult to set up an experimental realization of an integrable
system. This statement is corroborated by the fact that there is only
one kind of conductance experiment that shows a clear distinction
between integrable and chaotic underlying dynamics.\cite{Chang94}
However, this study, carried out by Chang and collaborators, dealt
with the weak localization peak in {\sl open} quantum dots, i.e., not
in the Coulomb blockade regime.

Looking for clear fingerprints of integrable dynamics in the
statistics of conductance in the Coulomb blockade regime is probably
not a well posed task. Integrable systems are not universal, implying
that each different confining potential will likely lead to very
different conductance fluctuations. In this section we pursue the
opposite strategy, asking if it is possible to tell from the
conductance experimental data if the underlying dynamics is chaotic or
not. To obtain some insight into this question we analyze two very
simple systems, namely, the rectangular billiard (a paradigm of
integrable motion) and the smooth stadium with $n=1$.

We model the coupling of the quantum dot to a given lead by a point
contact at the position $\bbox{r}_c$, which implies that the decay
width of the $p$th state is proportional to
$|\psi_p(\bbox{r}_c)|^2$. For billiards it is standard to solve the
single-particle problem with Neumann boundary conditions and to put
the point contacts at the boundary.\cite{Bruus94,Alhassid97} For
chaotic systems one obtains equivalent results choosing the point
contacts inside the dot, thus probing the ``bulk" of the
single-particle wave functions. In this section we investigate both
schemes.

For the rectangular billiard we can only proceed analytically in the
regime where $\Gamma \ll k_BT \ll \Delta$. The decay width of a state
with quantum numbers $(n,m)$ is given by $\Gamma = (4/L_xL_y) \sin^2
(k_n x)\sin^2(k_m y)$, up to a penetration factor. In the absence of
an external magnetic field, in the "bulk" point contact model,
$P(\Gamma)$ can be calculated through the expression

\begin{eqnarray}
\label{eq:squarefirst}
P_{\mbox{\scriptsize bulk}}(\Gamma) & = & \frac{1}{L_xL_y}\int_0^{L_x}
\!\!dx \int_0^{L_y} \!\!dy \; \delta \!\Biglb( \Gamma - \frac{4}{L_xL_y}
\sin^2(k_n x) \nonumber \\ & & \times \sin^2(k_m y) \Bigrb).
\end{eqnarray}
For point contacts, one immediately obtains $\langle \Gamma \rangle =
(L_xL_y)^{-1}$. The integrals in Eq.~(\ref{eq:squarefirst}) represent
an average over the position of the point contact. One could object
that this is not the kind of average taken in experiments. However, it
turns out that by reducing the integration in
Eq.~(\ref{eq:squarefirst}) to a single period in $x$ and $y$, the wave
function quantum numbers drop out of the integral and $P(\Gamma)$
becomes independent of $n$ and $m$. In this sense, we claim that the
average in Eq.~(\ref{eq:squarefirst}) is equivalent to an average over
a large number of wave functions with a fixed point contact position.

Performing the integral in Eq.~(\ref{eq:squarefirst}) and writing the
result in terms of $\langle \Gamma \rangle$, we arrive at
\begin{equation}
\label{eq:Pgamasquare}
P_{\mbox{\scriptsize bulk}}(\Gamma) = \frac{1}{\pi^2 \sqrt{
\langle\Gamma\rangle\,\Gamma}}
\int_{\sqrt{\Gamma/(4\langle\Gamma\rangle)}}^1 \frac{dz} {\sqrt{(1
-z^2)(z^2 - \Gamma/4\langle\Gamma\rangle)}},
\end{equation}
which can be written in a closed form in terms of a complete
elliptic integral of the first kind (not shown here). 

\begin{figure}
\setlength{\unitlength}{1mm}
\begin{picture}(100,80)(0,0)
\put(20,-35){\epsfxsize=12cm\epsfbox{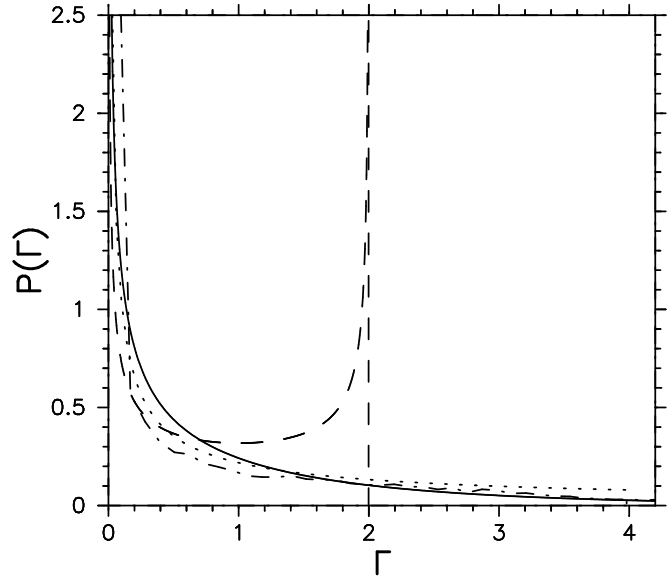}}
\end{picture}
\caption{$P(\Gamma)$ from RMT (solid line) compared with different
integrable systems: bulk rectangle (dotted), side rectangle (dashed),
and $n=1$ smooth stadium (dash-dotted). The distributions are scaled
to have $\langle \Gamma \rangle = 1$.}
\label{fig:Pgamasquare}
\end{figure}

In a similar manner, we can calculate the distribution of decay widths
$P(\Gamma)$ for the rectangular billiard with point contacts placed at
its sides, obtaining
\begin{equation}
P_{\mbox{\scriptsize side}}(\Gamma) = \frac{1}{\pi}
\frac{1}{\sqrt{\Gamma(2\langle\Gamma\rangle - \Gamma)}}.
\end{equation}
In Fig.~\ref{fig:Pgamasquare} we compare the analytical results of
$P(\Gamma)$ for the rectangle with the Porter-Thomas distribution,
which is the RMT prediction for the orthogonal symmetry. In addition,
we also display the numerical results for the smooth integrable
stadium potential ($n=1$).

The similarity between the bulk rectangle and the smooth stadium with
the Porter-Thomas distribution (in the absence of magnetic field) can
be understood by a simple argument. Any wave function has a discrete
number of points where its amplitude squared is a local maximum. It
has also nodal lines where the amplitude is zero. By picking
coordinates at random, the probability for obtaining a small amplitude
is much bigger than for a large amplitude. It has to be emphasized,
however, that neither the bulk rectangle nor the smooth stadium are in
perfect agreement with the Porter-Thomas distribution. Although
similar, there are differences that can be better visualized in a
log-log plot (not presented here), as shown before in the quantum
chaos literature (see, for example, Ref.~\onlinecite{Bruus94}).

\begin{figure}
\setlength{\unitlength}{1mm}
\begin{picture}(100,160)(0,0)
\put(20,50){\epsfxsize=12cm\epsfbox{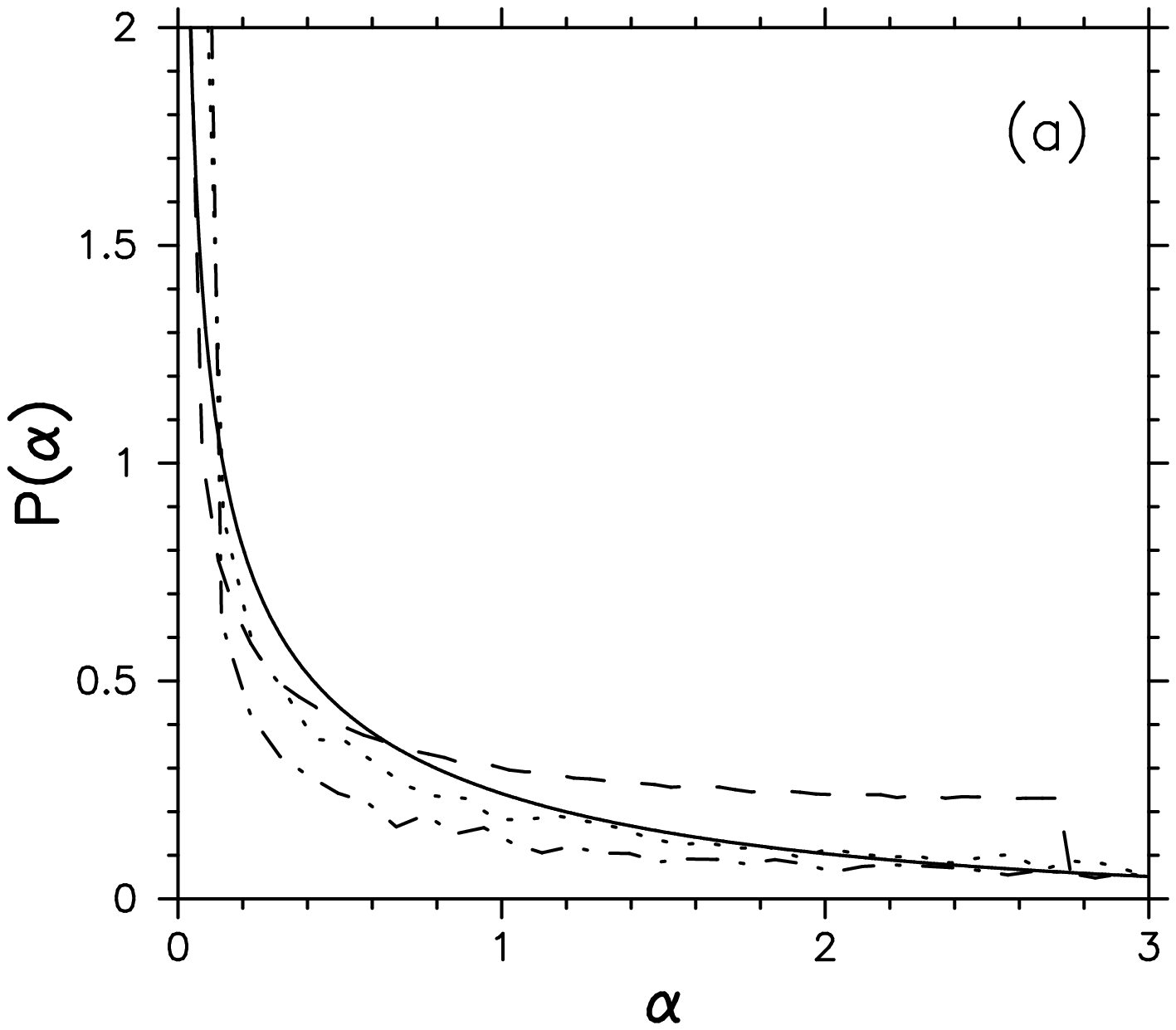}}
\put(20,-30){\epsfxsize=12cm\epsfbox{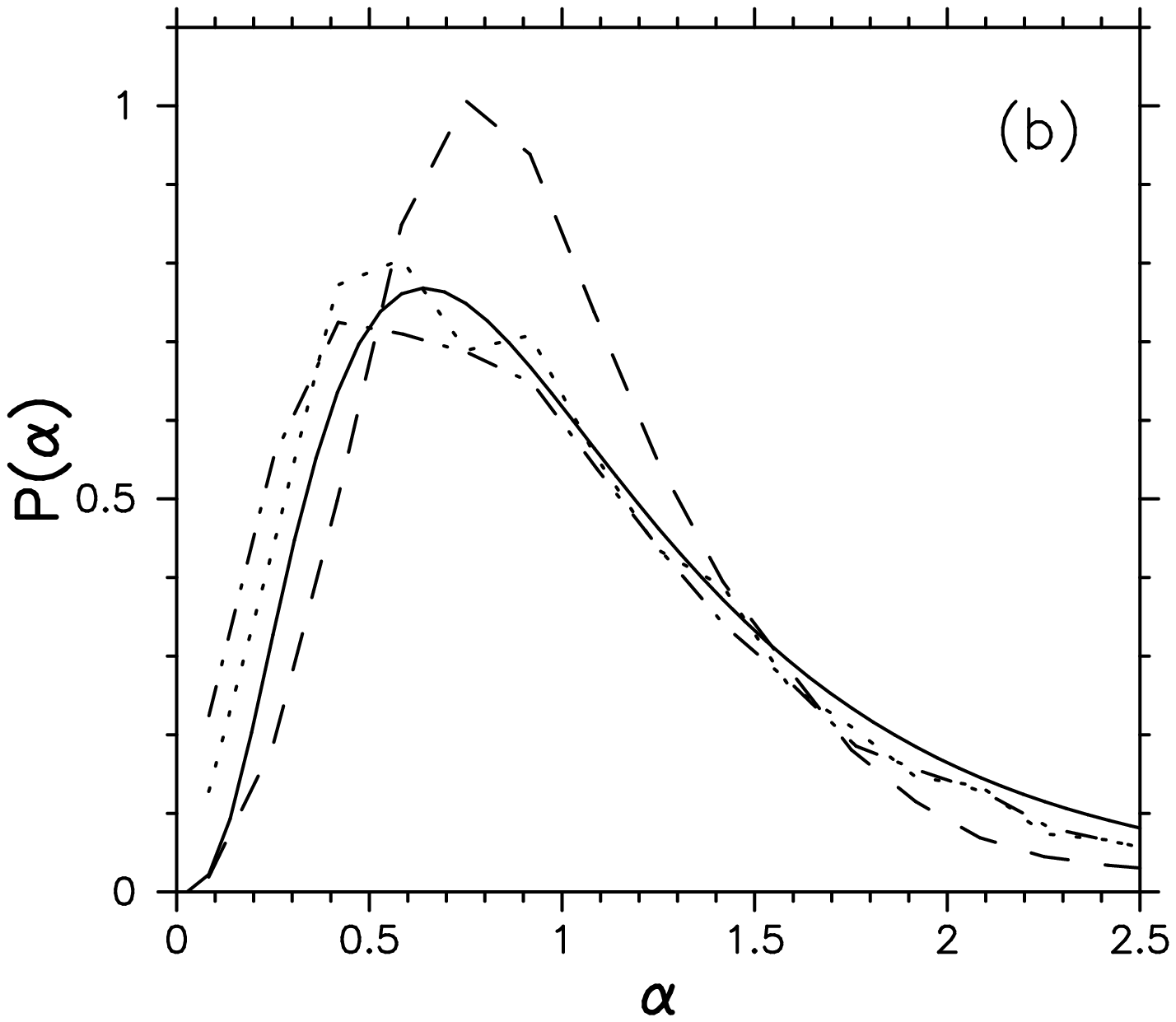}}
\end{picture}
\caption{$P(\alpha)$ from RMT (solid line) compared with different
integrable systems: bulk rectangle (dotted), side rectangle (dashed),
and $n=1$ smooth stadium (dash-dotted) for (a) $k_B T=0$ and (b)
$k_BT=1.0\Delta$. The distributions are scaled to have $\langle \alpha
\rangle = 1$.}
\label{fig:Pgintegravel}
\end{figure}

The most interesting aspect of this analysis appears when we treat the
distribution of conductance peak heights $P(\alpha)$. We find that
$P(\alpha)$ is relatively insensitive to $P(\Gamma)$, as illustrated
in Fig.~\ref{fig:Pgintegravel}. Analytical calculations are rather
difficult in this case, since, in general, we do not know how to take
into account correlations between the decay widths $\Gamma^l$ and
$\Gamma^r$ for integrable systems. Thus, the presented results were
obtained numerically. Notice that although the histograms are somewhat
different from the RMT result, they become indistinguishable when the
usual experimental uncertainties (5-10\%) are involved. (To construct
the histogram shown in Fig.~\ref{fig:Pgintegravel} we have used a very
large sample of conductance peak heights.) The apparent similarity
between the RMT prediction and the result for integrable systems
becomes even stronger if we take $k_BT =0.5\Delta$, as illustrated in
Fig.~\ref{fig:Pgintegravel}b.

From this analysis, we conclude that the universal aspect observed
experimentally for $P(\alpha)$ is more robust than initially
predicted. We believe that there is a large class of integrable
systems which lead to conductance peak distributions similar to
RMT. This result is not very intuitive, since for most observables
integrable systems show larger fluctuations than chaotic
ones.\cite{Alhassid89} We caution that our statement is {\sl not} that
{\sl any} integrable system will display $P(\alpha)$ similar to
Eqs.~(\ref{eq:PGOEg}) and (\ref{eq:PGUEg}). The literature already
offers us a counter example: The conductance fluctuations observed in
the model analyzed in Ref.~\onlinecite{Baltin98} are certainly not
compatible with RMT.


\section{Conclusions}
\label{sec:conclusions}

 
In this work we aimed to study quantitatively how Coulomb blockade
conductance peak statistics can be understood within the
single-particle plus constant interaction model. We showed that the
distribution of peak heights predicted by the RMT is very robust to
deformations in the dot effective potential. For $k_BT > 0.25 \Delta$
we found that there are exist integrable systems which give equivalent
results to RMT. The curves derived numerically from RMT reproduce
fairly the experimental results at low temperatures (for
$k_BT<0.5\Delta$), with increasing discrepancies as temperatures
rises.

We also showed that the peak height long-range correlations can be
understood in terms universal wave function correlations. The short
range correlations remain an open problem. Under certain
circumstances, we were able to reproduce peak sequences with visible
bunching of peak heights. Their correlation lengths depend on the
specific portion of the spectrum considered and is thus
nonuniversal. Unfortunately, such sequences are relatively rare, since
they rely on the existence of well-connected scarred states.

Starting from non-degenerate sequences of eigenenergies and tunneling
rates, one can construct spin-degenerate data sets by simply
duplicating the original sequences. In this case, at $k_BT/\Delta
\rightarrow 0$ the conductance peaks should appear in pairs of equal
heights. More importantly, paired peaks would have a similar response
to external perturbations, such as a magnetic field, even if their
heights are not quite identical. Experimentally, however, this
behavior is not observed (with the possible exception of Fig.~2 in
Ref.~\onlinecite{Patel98}). In view of this fact, we have decided for
restricting our analysis to nondegenerate single-particle levels.

It is noteworthy that the experiment by Chang and collaborators
\cite{Chang96} is the only one where peak bunching is not visually
strong. The main difference between the setup of
Ref.~\onlinecite{Chang96} and others is the sample, which in this case
has lower mobility. Disorder destroys the dynamical effects we
propose. One could then speculate that the lack of peak height
correlations in Ref.\onlinecite{Chang96}, and, perhaps, the departure
from universality seen in other experiments depend on the strength of
dynamical effects. From this perspective, our main conclusion is
negative, since, we were not able to identify a robust mechanism to
explain such phenomena based on this idea.

The hypothesis that the ground state and the excited states are
uncorrelated and follow RMT leads always to the suppression of small
conductance peak heights upon thermal averaging. Even the inclusion of
additional features, like deformations, does not change this
conclusion. In view of the experimental evidences of an even larger
suppression of the small peaks than predicted by RMT, we might
conclude that the ground state and the first excited states have
instead some correlation. The mechanism responsible for this could
also be responsible for the correlations between ground states with
consecutive number of electrons. Thus, peak bunching and the
persistence of small peaks could be manifestations of the same
mechanism. From our analysis, however, such mechanism seems to be
beyond the mean-field scenario. In conclusion, we believe that the
full quantitative understanding of these effects may lay outside the
single-particle approximation.


\acknowledgements 
 
We thank C. Marcus and S. Patel for making the experimental data on
the conductance peak sequences available to us. Discussions with
R. Jalabert, J. Richert, S. Tomsovic, and H. A. Weidenm\"uller were
greatly appreciated. One of us (C.H.L.) thanks the hospitality of the
Theoretical Physics Laboratory at the University of Strasbourg, where
this work was concluded. We gratefully acknowledge the financial
support of FAPERJ, CNPq, and PRONEX (Brazil).
 



\newpage

\newpage

\newpage

\newpage

\newpage

\newpage

\newpage

\newpage

\newpage

\newpage

\newpage

\end{document}